\documentclass[traditabstract]{aa}

\usepackage{graphicx}
\usepackage{txfonts}
\usepackage{hyperref}
\usepackage{xcolor}
\usepackage{cancel}
\graphicspath{{./}{figures/}}

\begin{document} 

\title{Exploring HNC and HCN line emission as probes of the protoplanetary disk temperature}

\titlerunning{HNC and HCN line emission in disks}

\author{Feng Long\inst{1,2} 
        \and Arthur D. Bosman\inst{3,4}
        \and Paolo Cazzoletti\inst{5}
        \and Ewine F. van Dishoeck\inst{3,5}
        \and Karin I. \"Oberg\inst{1}
        \and Stefano Facchini\inst{6}
        \and Marco Tazzari\inst{7}
        \and Viviana V. Guzm\'{a}n\inst{8}
        \and Leonardo Testi\inst{6}  }

\institute{Harvard-Smithsonian Center for Astrophysics, 60 Garden Street, Cambridge, MA 02138, USA \\ \email{feng.long@cfa.harvard.edu}
\and Kavli Institute for Astronomy and Astrophysics, Peking University, Beijing 100871, China
\and Leiden Observatory, Leiden University, Niels Bohrweg 2, 2333 CA Leiden, The Netherlands 
\and Department of Astronomy, University of Michigan, 323 West Hall, 1085 S. University Avenue, Ann Arbor, MI 48109, USA 
\and Max-Planck-Institut f\"{u}r Extraterrestrische Physik, Giessenbachstrasse 1, 85748, Garching, Germany 
\and European Southern Observatory, Karl-Schwarzschild-Str. 2, D-85748 Garching bei M\"{u}nchen, Germany 
\and Institute of Astronomy, University of Cambridge, Madingley Road, Cambridge CB3 0HA, UK
\and Instituto de Astrof\'isica, Pontf\'ificia Universidad Cat\'olica de Chile, Av. Vicu\~na Mackenna 4860, 7820436 Macul, Santiago, Chile
}

\date{Received xx; accepted xx}

\abstract
{The distributions and abundances of molecules in protoplanetary disks are powerful tracers of the physical and chemical disk structures. The abundance ratios of HCN and its isomer HNC are known to be sensitive to gas temperature. Their line ratios might therefore offer a unique opportunity to probe the properties of the emitting gas.}
{We investigate the HNC and HCN line emission in disks at (sub-)millimeter wavelengths and explore their potential utility for probing disk temperature and other disk properties.}
{Using the 2D thermochemical code DALI, we ran a set of disk models accounting for different stellar properties and radial and vertical disk structures, with an updated chemical network for the nitrogen chemistry. These modeling results were then compared with observations, including new observations obtained with the Atacama Large Millimeter/submillimeter Array (ALMA) of HNC $J=3-2$ for the TW Hya disk and HNC $J=1-0$ for 29 disks in Lupus.}
{Similar to CN, HCN and HNC have brighter line emission in models with larger disk flaring angles and higher UV fluxes. HNC and HCN are predicted to be abundant in the warm surface layer and outer midplane region, which results in ring-shaped emission patterns. However, the precise emitting regions and emission morphology depend on the probed transition, as well as on other parameters such as C and O abundances. The modeled HNC-to-HCN line intensity ratio increases from $<0.1$ in the inner disk to up to 0.8 in the outer disk regions, which can be explained by efficient HNC destruction at high temperatures. Disk-integrated HNC line fluxes from current scarce observations and its radial distribution in the TW Hya disk are broadly consistent with our model predictions. }
{The HNC-to-HCN flux ratio robustly increases with radius (decreasing temperature), but its use as a chemical thermometer in disks is affected by other factors, including UV flux and C and O abundances. High-spatial resolution ALMA disk observations of HNC and HCN that can locate the emitting layers would have the great potential to constrain both the disk thermal and UV radiation structures, and also to verify our understanding of the nitrogen chemistry.} 

\keywords{astrochemistry --- protoplanetary disks --- stars: pre-main sequence ---  planet formation
               }

\maketitle
%

\section{Introduction} \label{sec:intro}
Gas-rich protoplanetary disks around young stars provide raw materials for the assembly of planets. The final properties of the planetary systems are therefore largely determined by the physical and chemical conditions of disks. Emission lines of small molecules and radicals are excellent probes of disk ionization, density, and thermal structures \citep[e.g.,][]{vanZadelhoff2001, Teague2016}. In addition to the simple CO molecule, a variety of molecules with strong enough line emission (including CN, CS, HCN, HNC, C$_2$H, c-C$_3$H$_{2}$ ,H$_2$CO, CH$_3$OH, and CH$_3$CN) at (sub-)millimeter wavelengths have been detected in disks, tracing the disk layers where they originate \citep[e.g.,][]{Dutrey1997, Kastner1997,Kastner2014,Kastner2018,Thi2004, Oberg2010,Punzi2015,Walsh2016,Bergner2018,Bergner2019, Pegues2020, Loomis2020}.

With the advent of high-resolution imaging, investigations of the detailed spatial distributions of disk materials become possible. Strikingly, ring-like structures of dust grain distributions emerge in most high-resolution continuum maps \citep[e.g.,][]{Andrews2018, Long2018, vandermarel2019}. In the so-called transition disks, where the inner regions are depleted of millimeter-sized grains, gas cavities are observed with CO isotopolog lines as well \citep{vanderMarel2016}. Ring-shaped emission has also been found in other molecules, but these molecular rings are not always associated with dust rings: $\rm N{_2}H^{+}$ and $\rm DCO^{+}$ emission rings trace regions near the CO snow line, where the disk temperature drops for CO gas to freeze out onto grains \citep{Qi2013,Mathews2013,Oberg2015}\footnote{The emergence of an additional exterior $\rm DCO^{+}$ emission ring could be explained by the nonthermal desorption of CO ice \citep{Oberg2015}.}; CN emission rings are strongly linked to the local UV field \citep{Cazzoletti2018}; the formation of hydrocarbon rings (e.g., $\rm C{_2}H$ and $\rm C{_3}H_{2}$) requires not only a strong UV field, but also an elevated C/O \citep{Bergin2016, Miotello2019}. These examples highlight the possibility of using chemical structures as probes of disk conditions. Fully realizing the potential utility of molecules as tracers of different disk properties would  have profound implications for studies of protoplanetary disks and planet formation.

The simple organic molecule hydrogen cyanide (HCN) and its isomer HNC are of particular interests for this purpose. Isomers usually have connected formation and destruction pathways regulated by physical characteristics, and their line ratios therefore offer a unique opportunity to probe the properties of the emitting gas.  HNC and HCN have been detected in a variety of astrophysical environments, including different phases of star formation \citep{Irvine1984, Schilke1992, Hirota1998,Padovani2011}, proto-brown dwarfs \citep{Riaz2018}, protoplanetary disks \citep[][]{Dutrey1997,Graninger2015}, and also planetary nebulae \citep[][]{Schmidt2017,Bublitz2019}.
These observations often reveal a temperature-dependent HNC-to-HCN line ratio, for instance, with near-unity ratios found in cold dark clouds, and very low ratios detected toward warm protostellar cores \citep[][]{Schilke1992, Padovani2011} and hot ultracompact H II regions related to massive star formation \citep[][]{Jin2015}. A decrease in HNC-to-HCN line ratio with UV luminosity has been found in planetary nebulae \citep{Bublitz2019}, but this could also be (partially) explained by the temperature effect, with UV photons heating the gas.
\citet{Hacar2020} have recently established the HNC-to-HCN line intensity ratio as a chemical thermometer for the cold interstellar medium (ISM) based on observations towards the integral shape filament of Orion. Temperature is a fundamental parameter in protoplanetary disks. It sets the locations of molecular ice lines \citep{Qi2019}, regulates the physical and chemical evolution of the disk \citep[e.g.,][]{Kenyon1995, Dutrey2014}, and is vital for determining the disk mass \citep[e.g.,][]{Trapman2017}. While the HCN lines are usually bright and readily detected in disks \citep[][]{Oberg2011, Chapillon2012, Guzman2015}, the fainter HNC lines are rarely targeted \citep[][]{Dutrey1997}.  \citet{Graninger2015} presented the first spatially resolved observations of HNC lines in disks with a ring-shaped emission structure, which they interpreted as temperature-regulated HNC destruction.  This result therefore demonstrates that spatially resolved HNC and HCN observations may be employed to map the disk temperature structure.

Because the excitation conditions for HNC and HCN are similar, the line intensity ratio would reflect their relative abundances and should be regulated by chemical reactions. The two isomers are mainly produced by the dissociative recombination of HCNH$^+$, 
\begin{eqnarray}
{\rm HCNH}^+ \ +\ {\rm e}^-  \ \to\ &\  {\rm HCN} \ +\ {\rm H}\, \\ 
{\rm HCNH}^+ \ +\ {\rm e}^-  \ \to\ &\  {\rm HNC} \ +\ {\rm H}\,\label{rx:hcnh+} 
,\end{eqnarray}
which can be generated from pathways involving NH$_3$, atomic N, and N$_2$ \citep{Loison2014}. This reaction has an approximately equal branching ratio, therefore the abundance differences between HCN and HNC are largely determined by the main selective destruction pathways of HNC,
\begin{eqnarray}
{\rm HNC} \ +\ {\rm C}  \ \to\ &\  {\rm HCN} \ +\ {\rm C}\, \label{rx:hnc/c} \\
{\rm HNC} \ +\ {\rm H}  \ \to\ &\  {\rm HCN} \ +\ {\rm H}\, \label{rx:hnc/h} \\ 
{\rm HNC} \ +\ {\rm O}  \ \to\ &\  {\rm CO} \ +\ {\rm NH}\,. \label{rx:hnc/o}
\end{eqnarray}
The rate coefficient of reaction~\ref{rx:hnc/c} is constant with temperature \citep{Loison2014}, while reactions~\ref{rx:hnc/h} and \ref{rx:hnc/o} possess activation barriers of 200\,K \citep{Graninger2014} and 20\,K \citep{Hacar2020}, respectively, to proceed. Reaction~\ref{rx:hnc/h} is already active at $\sim$20\,K, but the reaction rate is orders of magnitude lower than at 200\,K. Because the majority of disk areas are warmer than 20\,K, the temperature dependence of the HNC-to-HCN line ratio is  primarily controlled by reaction~\ref{rx:hnc/h}. Although the line ratio also depends on elemental (e.g., carbon or oxygen) abundances, a high HNC-to-HCN line ratio would indicate a cold region of the disk, however.

We present here the first exploration of HNC and HCN modeling in protoplanetary disks, employing the 2D thermochemical code called "dust and lines" (DALI, \citealt{Bruderer2012,Bruderer2013}). Our goal is to investigate the HNC and HCN emission in protoplanetary disks based on our knowledge of cyanide chemistry and to explore the potential usability of HNC and HCN as tracers of the disk physical and chemical conditions, especially as a disk thermometer. Section~\ref{sec:model} describes the physical framework and the chemical network. In Section~\ref{sec:results} we present the results of molecular abundances and line emission from our models, and we compare the results with currently available disk observations, including new ALMA observations of HNC $J=3-2$ in TW Hya disk and HNC $J=1-0$ for a sample of Lupus disks.  Section~\ref{sec:diss} discusses caveats and possible solutions in using HNC-to-HCN ratio as disk thermometer. We summarize our main findings in Section~\ref{sec:sum}.

\section{Physical and chemical models} \label{sec:model}
To model the HNC and HCN emission in protoplanetary disks, we employed the 2D thermochemical code DALI (\citealt{Bruderer2012,Bruderer2013}), which has been widely used to model gas emission in disks \citep{Miotello2014,vanderMarel2015,Kama2016,Facchini2017,Trapman2017,Trapman2019}.
This code includes calculations of radiative transfer, chemistry, thermal balance, and ray-tracing. Provided with the dust and gas density structures, DALI first solves the dust continuum radiative transfer based on a Monte Carlo method to obtain the local UV flux and dust temperature $T_{\rm dust}$ at each position of the disk.  $T_{\rm dust}$ is used as an initial guess for the gas temperature $T_{\rm gas}$, and the abundances for chemical species are calculated with a chemical network simulation and are then fed into the non-LTE excitation calculation to solve $T_{\rm gas}$ through the heating-cooling balance. Because both the abundance and excitation calculations depend on $T_{\rm gas}$, the final $T_{\rm gas}$ is obtained by iterating the above steps until a self-consistent solution is reached. Spectral line cubes are then created with a raytracer for a given source distance and disk inclination. In this section, we describe the physical properties for the star and disk system, as well as the chemical network used in our models.

\subsection{Physical framework}
The main input for physical parameters is the disk density structure. We adopt the gas surface density profile as for a viscous evolving accretion disk (viscosity $\nu \propto R^{\gamma}$, \citealt{Lynden-Bell1974, Hartmann1998}), 
\begin{equation}\label{eq:dens}
\Sigma_{\rm gas}(R)=\Sigma_{\rm c}\bigg(\frac{R}{R_{\rm c}}\bigg)^{-\gamma}\exp \Bigg[-\bigg(\frac{R}{R_{\rm c}}\bigg)^{2-\gamma}\Bigg],
\end{equation}
where $\Sigma_{\rm c}$ is the surface density at the characteristic radius $R_{\rm c}$ and is set to yield the required total disk mass. The power-law index $\gamma$ is taken as 1 \citep{Hughes2008,Andrews2011} and assumed to not change with time. The gas density in the vertical direction follows a Gaussian distribution, with a scale-height angle of $h=h_{c}(R/R_{c})^{\psi}$. The physical scale height is thus given as $H \sim Rh$. 
The dust surface density is scaled from the gas surface density assuming a typical gas-to-dust ratio of 100. Because larger grains are more settled toward the midplane, we consider two populations of grains, following \citet{DAlessio2006}: a small-grain population (0.005-1\,$\mu$m) with the same scale height $h$ as the gas, and a large-grain population (1-1000\,$\mu$m) with a reduced scale height of ${\chi}h$, where ${\chi}=0.2$. The fraction of dust surface density distributed to the large grains is described by $f_{\rm large}$ such that $\Sigma_{\rm dust}=f_{\rm large}\Sigma_{\rm large}+(1-f_{\rm large})\Sigma_{\rm small}$. We have tested models in which we varied the $f_{\rm large}$ parameter between 0.9 and 0.99, but found  no significant differences in line emission of cyanides \citep{Cazzoletti2018}. We took $f_{\rm large}=0.99$ in our models because large grains dominate the dust mass budget.

High-energy radiation from the central star and interstellar medium affects the disk chemistry. The stellar spectrum is modeled as a blackbody for a given effective temperature $T_{\rm eff}$ and stellar luminosity $L_\ast$. The FUV spectrum (6--13.6\,eV) is particularly important for the chemistry of cyanides, therefore we considered two types of stars: a T Tauri star with $T_{\rm eff}$=4000\,K and $L_\ast$=1\,$L_\odot$, and a Herbig Ae star with $T_{\rm eff}$=10000\,K and $L_\ast$=10\,$L_\odot$, which we refer to as T Tauri and Herbig models throughout the paper. 
Additional UV flux from accretion can also be included for T Tauri stars, modeled as a blackbody emitting at 10000\,K on to a 1$M_{\odot}$ and 1.5$R_{\odot}$ star, with the total accretion luminosity controlled by the mass accretion rate $\dot{M}_{\rm acc}$. We took the typical accretion rate for young T Tauri stars of $10^{-8}\,\rm M_{\odot}\,yr^{-1}$ \citep{Hartmann2016, Manara2016} as the default in our model and varied the value to explore the effects of UV radiation on cyanide chemistry (see Figure~1 of \citealt{Visser2018} for the UV flux change with varying accretion rates and Figure~11 of \citealt{Cazzoletti2018} for the consequences on CN emission). We set an interstellar UV flux of $G_0$, as defined in \citet{Draine1978}, of  $\rm \sim2.7\times10^{-3}\,erg\,s^{-1}\,cm^{-2}$ between 6 and 13.6\,eV range, and a cosmic-ray ionization rate of $\rm 5\times10^{-17}\,s^{-1}$ per $\rm H_2$. The X-ray spectrum was taken as thermal radiation at $7\times10^7$\,K between 1 and 100\,keV, with the X-ray luminosity of 10$^{30}$\,erg\,s$^{-1}$. The same prescription was used for the T Tauri and Herbig models. While abundances of some molecules or molecular ions such as HCO$^{+}$ may be sensitive to the choice of X-ray luminosity or temperature, parameter studies have shown that most species, including HCN, HNC, and their ratio, have only small variations over the wide range of X-ray values that is applicable to T Tauri and Herbig stars \citep{Stauber2005, Bruderer2009, Bruderer2013, Cleeves2017}.
A small grid of models was performed by varying disk mass, disk vertical height, and flaring for the T Tauri and Herbig models to represent a wide range of physical and chemical conditions. The disk and stellar parameters used in the models are listed in Table~\ref{tab:parameters}.

\begin{table}[!tbh]
\caption{Disk and stellar parameters for the models.}
\label{tab:parameters}
\centering
\begin{tabular}{ll}
\hline\hline
Parameter        & Range \\
\hline
\emph{Chemistry} \\
$\rm [C]/[H]$ & 1$\times 10^{-4}$\\
$\rm [O]/[H]$ & 3.5$\times 10^{-4}$\\
$\rm [N]/[H]$ & 1.6$\times 10^{-5}$\\
\\
\emph{Physical structure}& \\
$\gamma$ & 1\\
$\psi$ & 0.1, 0.2, 0.3 \\
$h_{\rm c}$ & 0.1, 0.2  rad \\
$R_{\rm c}$ & 60 au \\
$M_{\rm gas}$ & $10^{-5} , 10^{-4} , 10^{-3}, 10^{-2}, 10^{-1} M_\odot$ \\
$f_{\rm large}$ & 0.99\\
$\chi$ & 0.2 \\
\\
\emph{Stellar spectrum}&\\
$T_{\rm eff}$ & 4000 K +  UV ($\dot M=10^{-8}\,\rm M_\odot/yr$),\\
&10000 K\\
$L_{\rm bol}$ & 1, 10 $L_{\odot}$\\
$L_{\rm X}$ & $\rm 10^{30}\, erg\,s^{-1}$\\
\\
\emph{Dust properties}&\\
Dust & 0.005-1 $\mu$m (small)\\
& 1-1000 $\mu$m (large)\\
\emph{Other parameters}&\\
Cosmic-ray ionization& \\
rate per H$_2$ & $5\times10^{-17}{\rm s}^{-1}$ \\
 External UV flux & $G_0$\\
\hline
\end{tabular}
Note: Chemistry -- initial abundances for C, O, and N. The effect of different elemental abundances for the HNC and HCN abundances is also explored in Section~\ref{sec:C/O}. Physical structure -- power-law index ($\gamma$) for surface density, power-law index ($\psi$) for the scale height, scale height ($h_c$) at critical radius ($R_c$), disk mass, fraction of large-grain population, and its settling parameter ($\chi$).
\end{table}

\subsection{Chemical network}
Our models used the new reduced network presented in \citet{Visser2018} for CNO chemistry, which is based on the network of \citet{Bruderer2012} and includes the reactions for cyanides discussed in \citet{Loison2014}. The entire chemical model is the same as was used for the investigation of CN emission in disks \citep{Cazzoletti2018}, except for two changes made relating to the destruction of HNC: (1) we used an updated reaction rate coefficient of HNC+H based on \citet{Graninger2014} with an energy barrier of 200\,K, and (2) we included the reaction of HNC+O with an energy barrier of 20\,K based on the recent study of the HNC-to-HCN ratio in molecular clouds by \citet{Hacar2020}. The updated network has negligible effects on the CN emission. Reactions involving isotopologs were not considered. The same photodissociation rate was used for HCN and HNC; the actural HNC photodissociation rate may be a factor of 2 higher than that of HCN \citep{Aguado2017}.

\begin{figure}[!th]
\centering
    \includegraphics[width=0.45\linewidth]{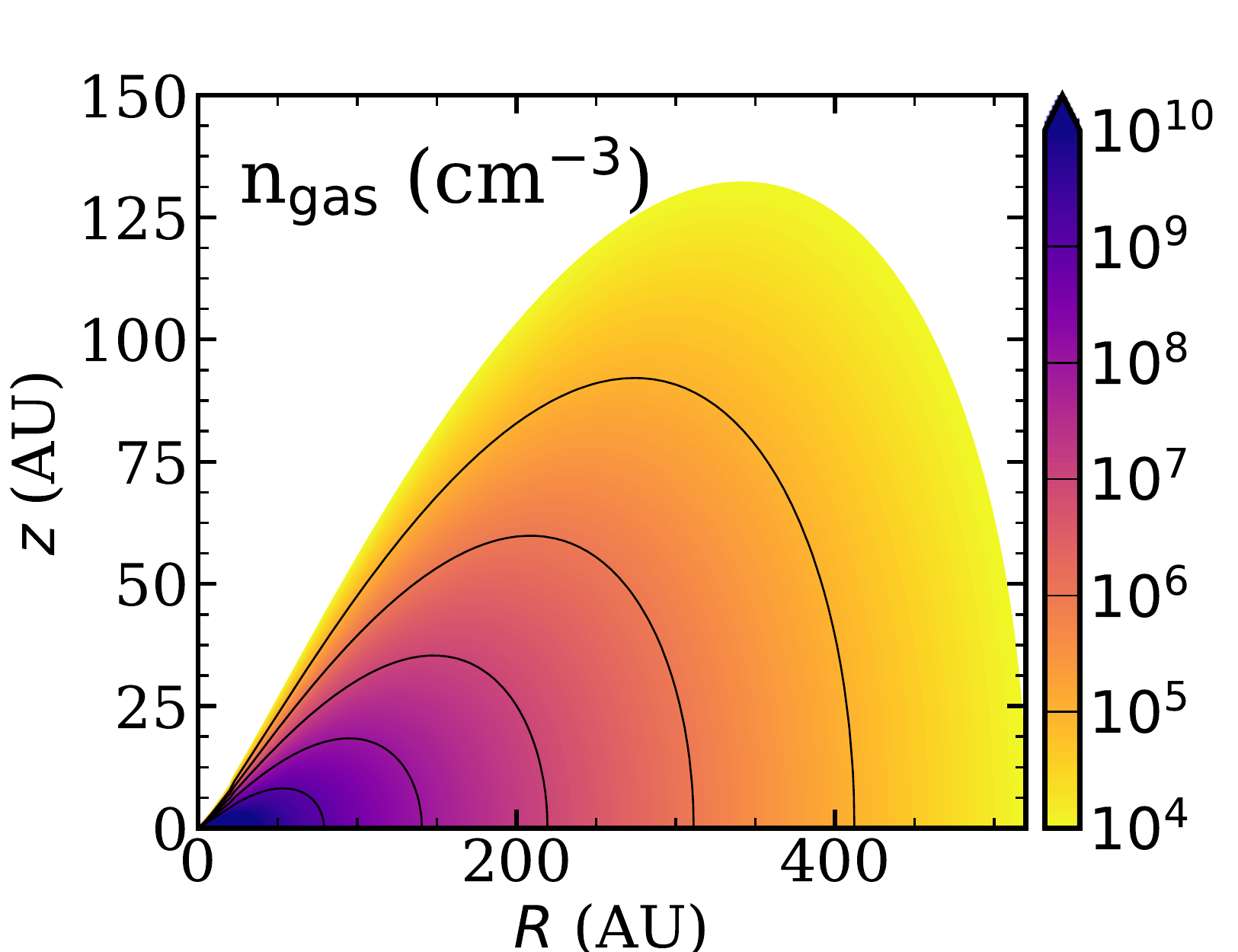}
    \includegraphics[width=0.45\linewidth]{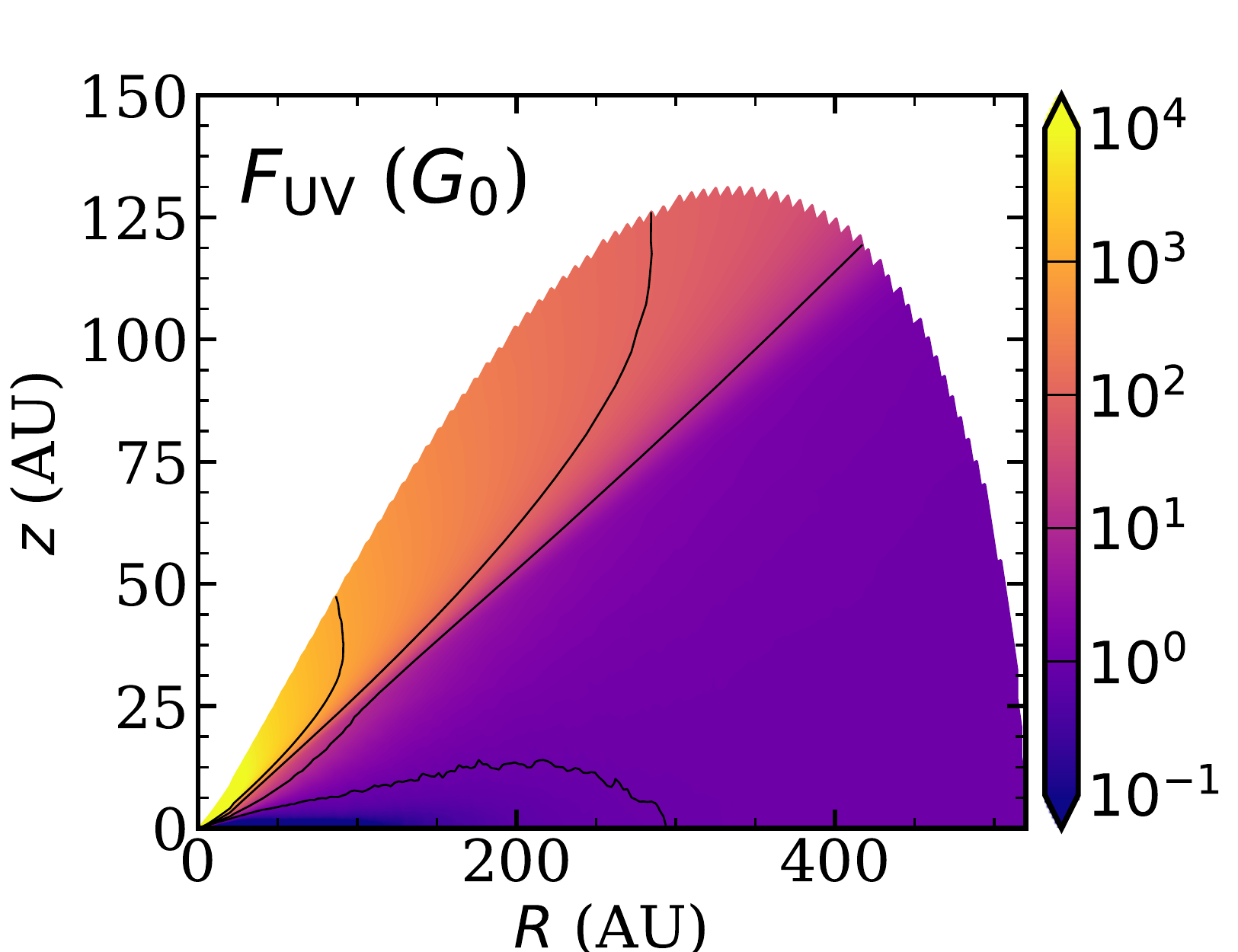} \\
    \includegraphics[width=0.45\linewidth]{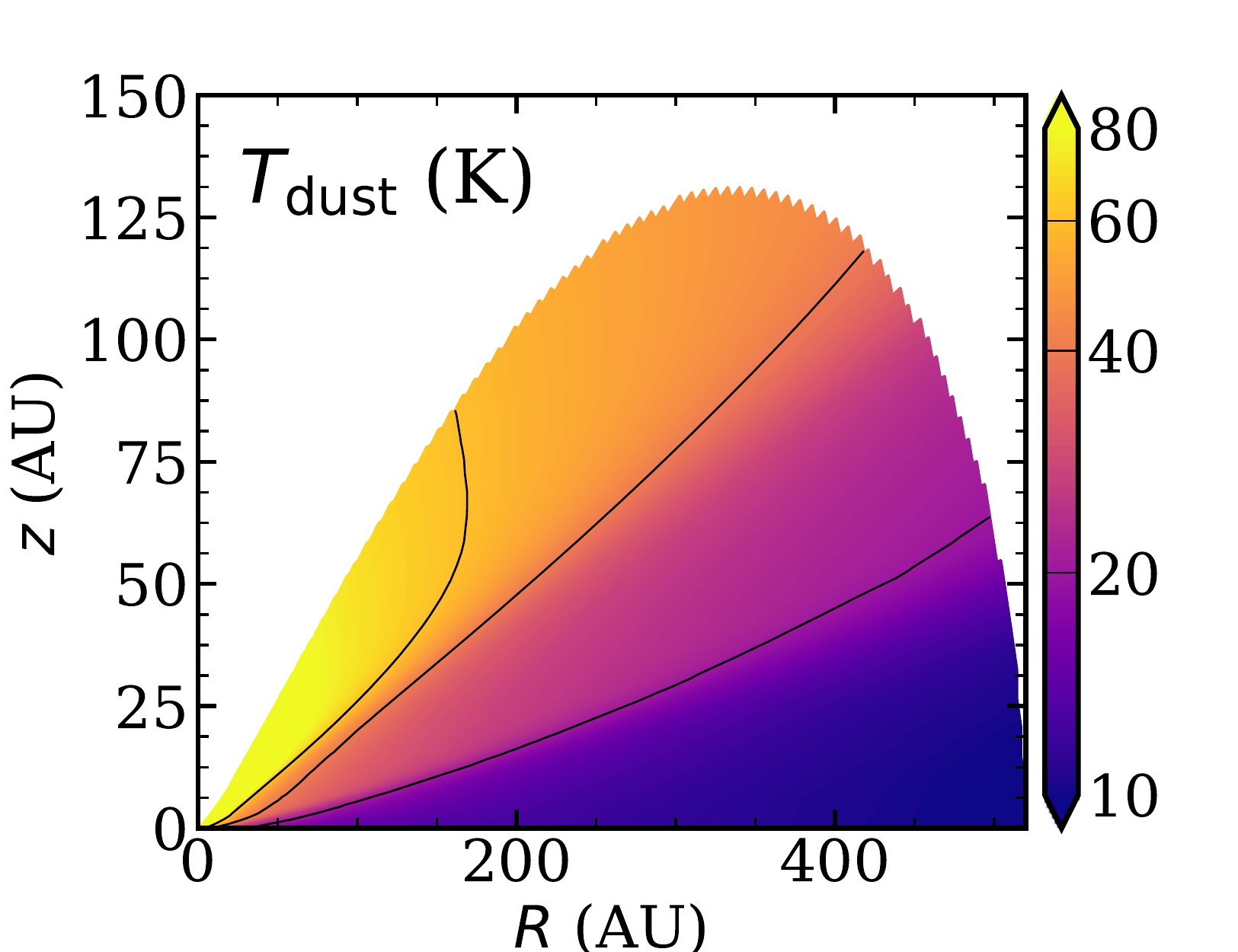}
    \includegraphics[width=0.45\linewidth]{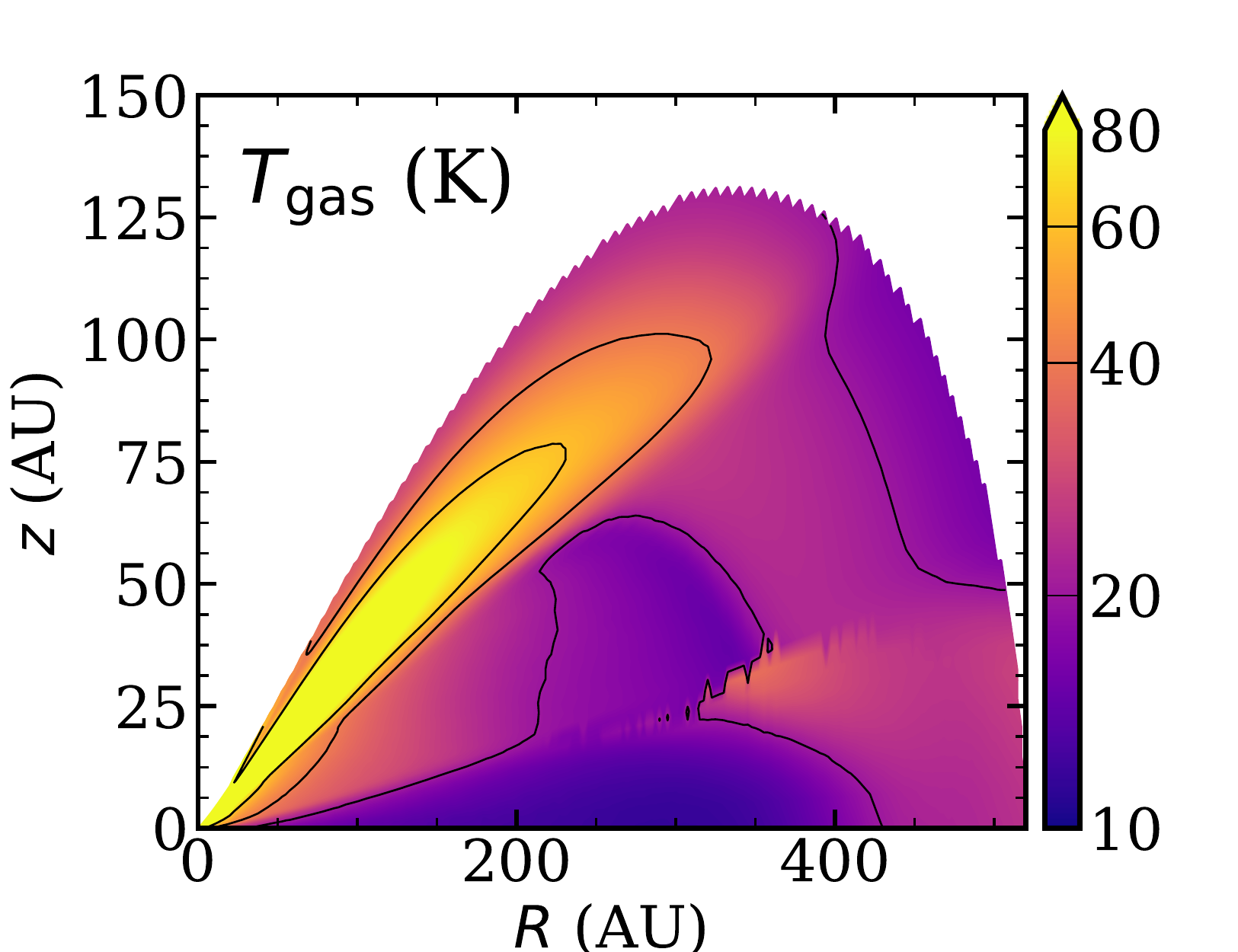} \\
\caption{Two-dimensional profiles of gas density, UV flux, dust temperature, and gas temperature for our fiducial model: a $10^{-2}\,\rm M_{\sun}$ disk with $\psi$=0.2, $h_c$=0.1 surrounding a T Tauri star. The UV flux is plotted in units of the mean interstellar radiation field ($G_0$, \citealt{Draine1978}). Temperature only shows a limited range to highlight the cold outer disk region. \label{fig:model}}
\end{figure}

The network contains  gas-phase and grain-surface reactions, including neutral-neutral and ion-molecular chemistry, hydrogenation of simple species on ices, photodissociation and photoionization, freeze-out and desorption, X-ray and cosmic-ray induced reactions, and reactions with vibrationally excited $\rm H_2$. The details of these processes are elaborated in \citet{Bruderer2012} and \citet{Visser2018}. We used binding energies to grains of 1600\,K for CN and 2050\,K for HNC and HCN, based on the values from the Kinetic Database for Astrochemistry (KIDA; \citealt{Wakelam2012}). The chemistry was run in steady-state mode with initial ISM abundances for CNO as listed in Table~\ref{tab:parameters}, thus with the ISM C/O. Given the uncertainties in individual rate coefficients and physical structure, our model abundances and column densities have uncertainties of at least a factor of a few. The trends in abundances and line fluxes with model parameters are more robust.

\begin{figure*}[!th]
\centering
    \includegraphics[width=0.24\linewidth]{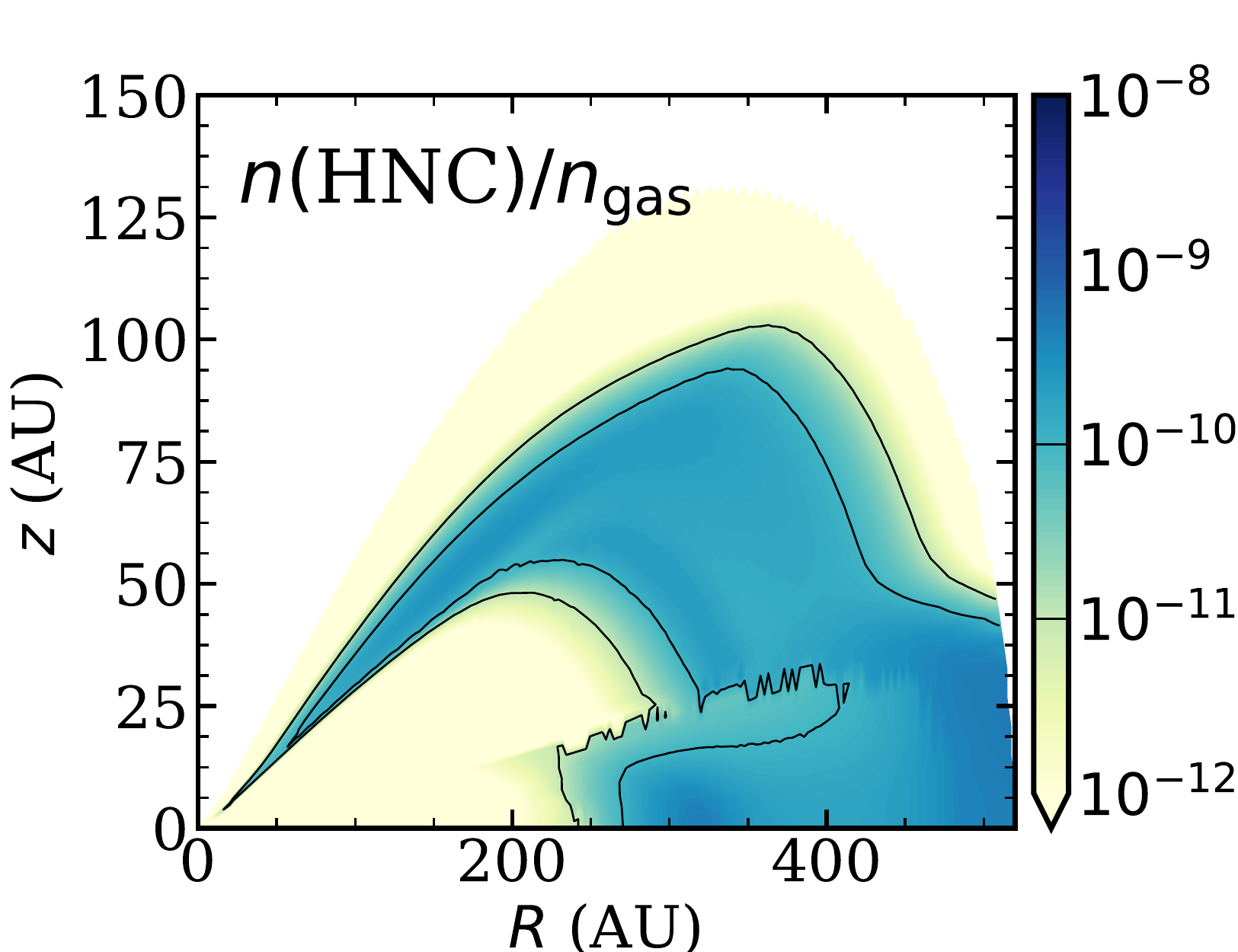}
    \includegraphics[width=0.24\linewidth]{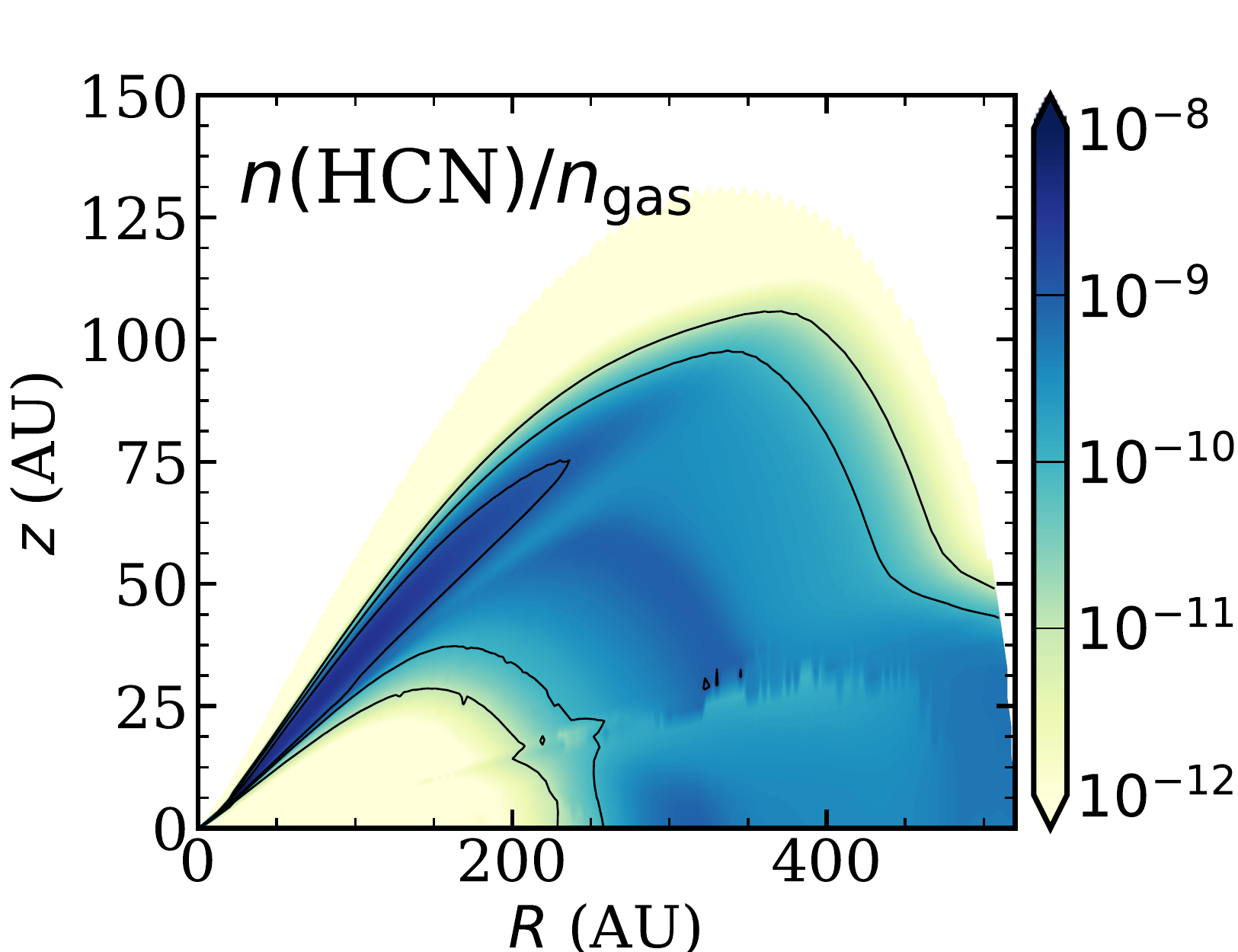}
    \includegraphics[width=0.24\linewidth]{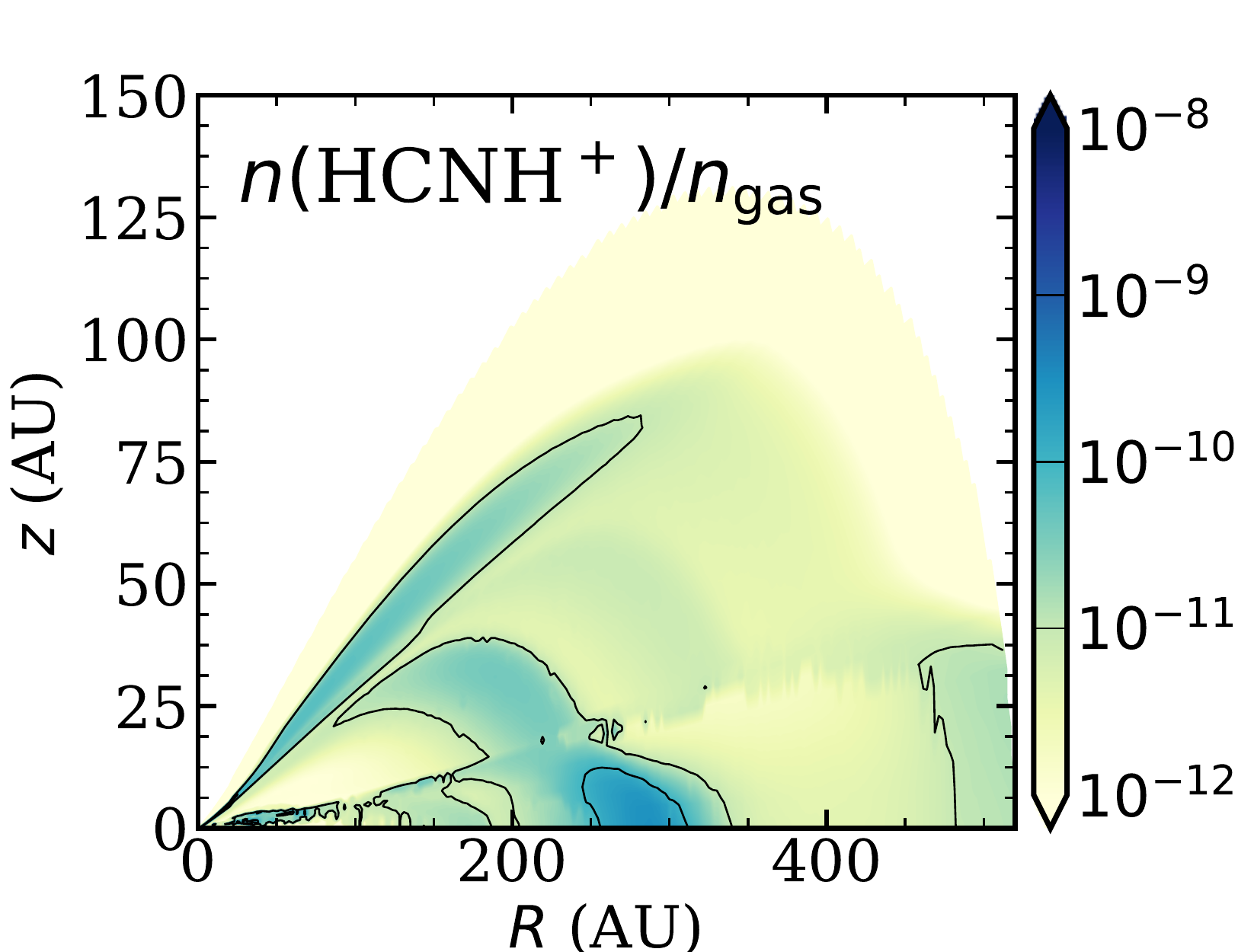}
    \includegraphics[width=0.24\linewidth]{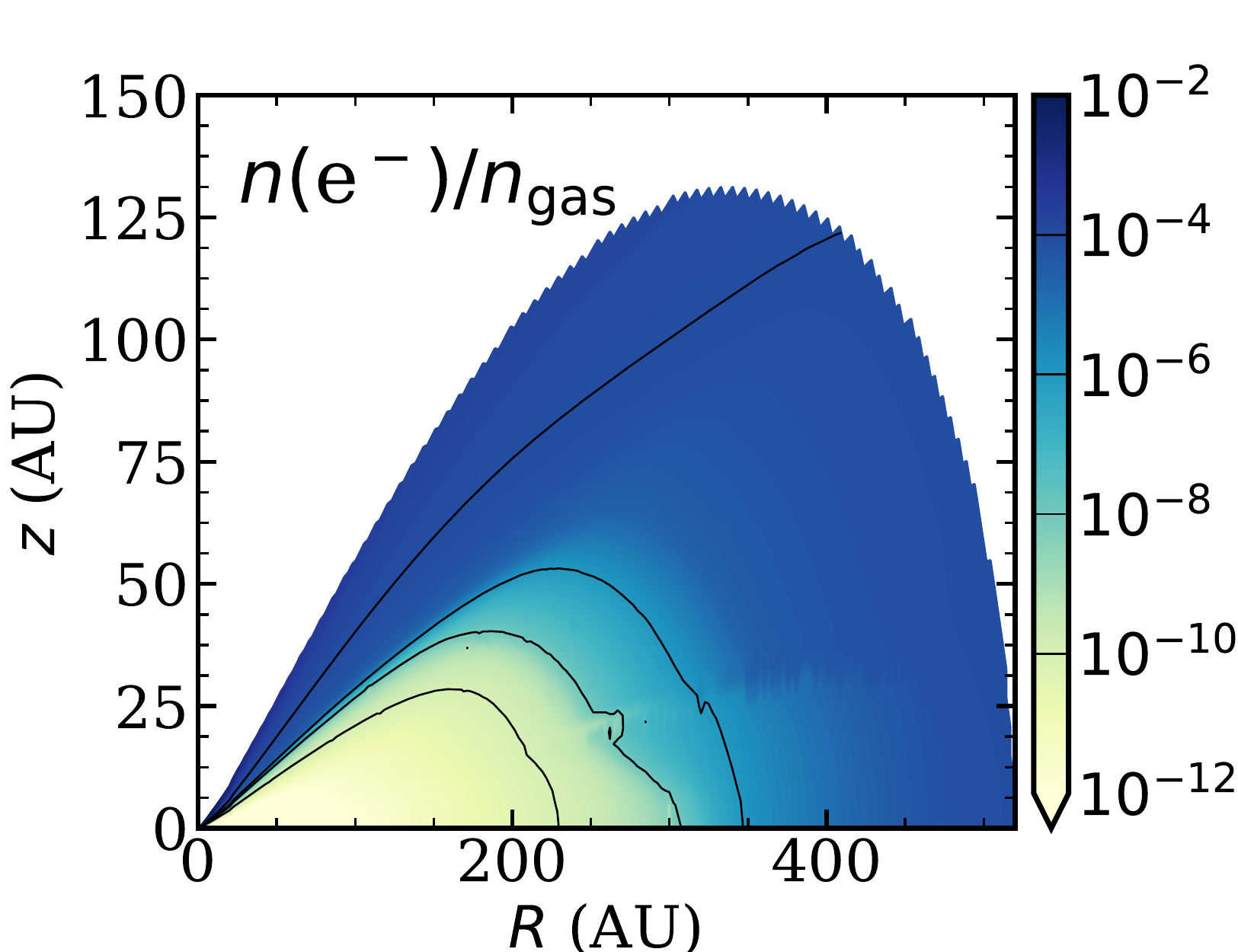}  \\
    \includegraphics[width=0.24\linewidth]{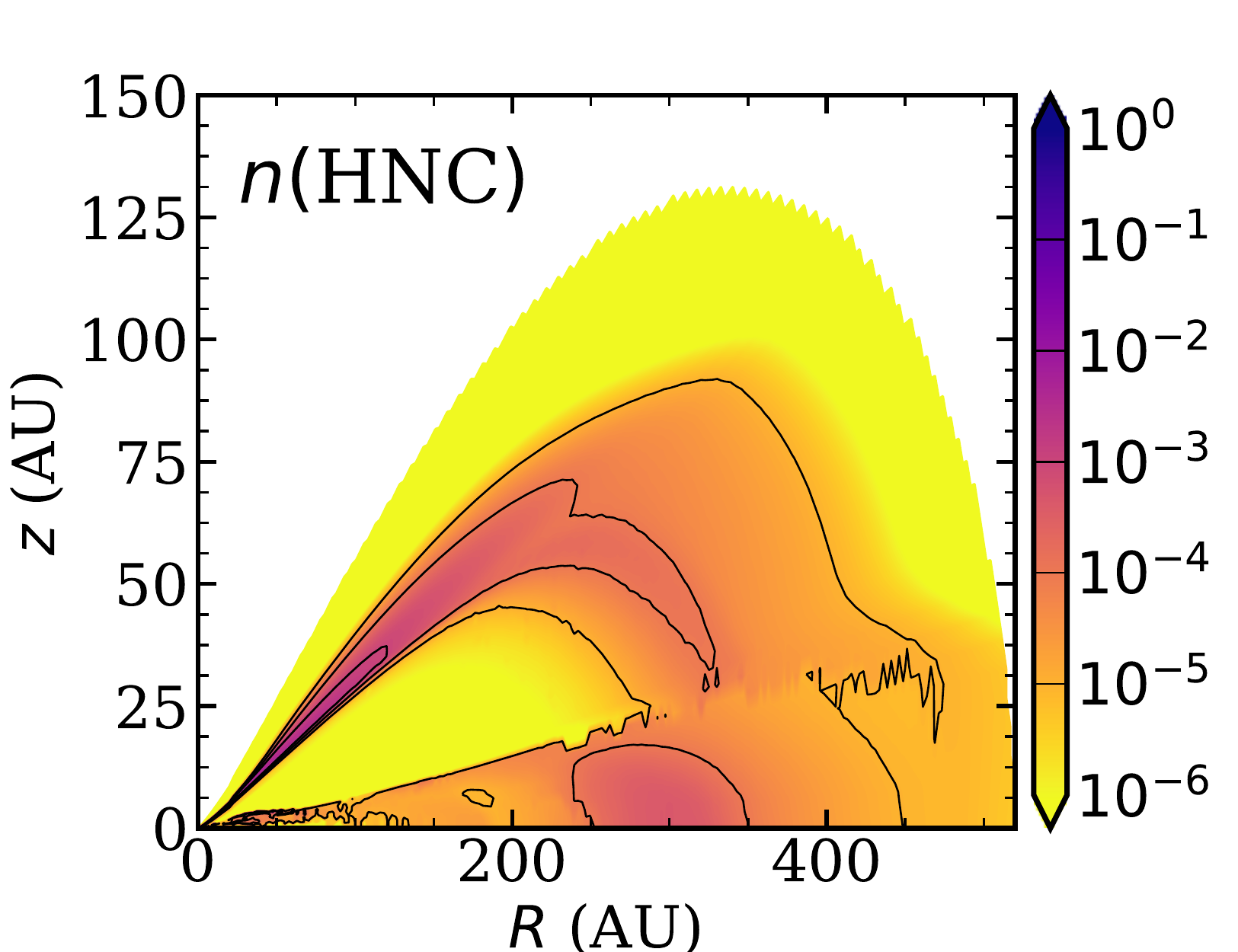}
    \includegraphics[width=0.24\linewidth]{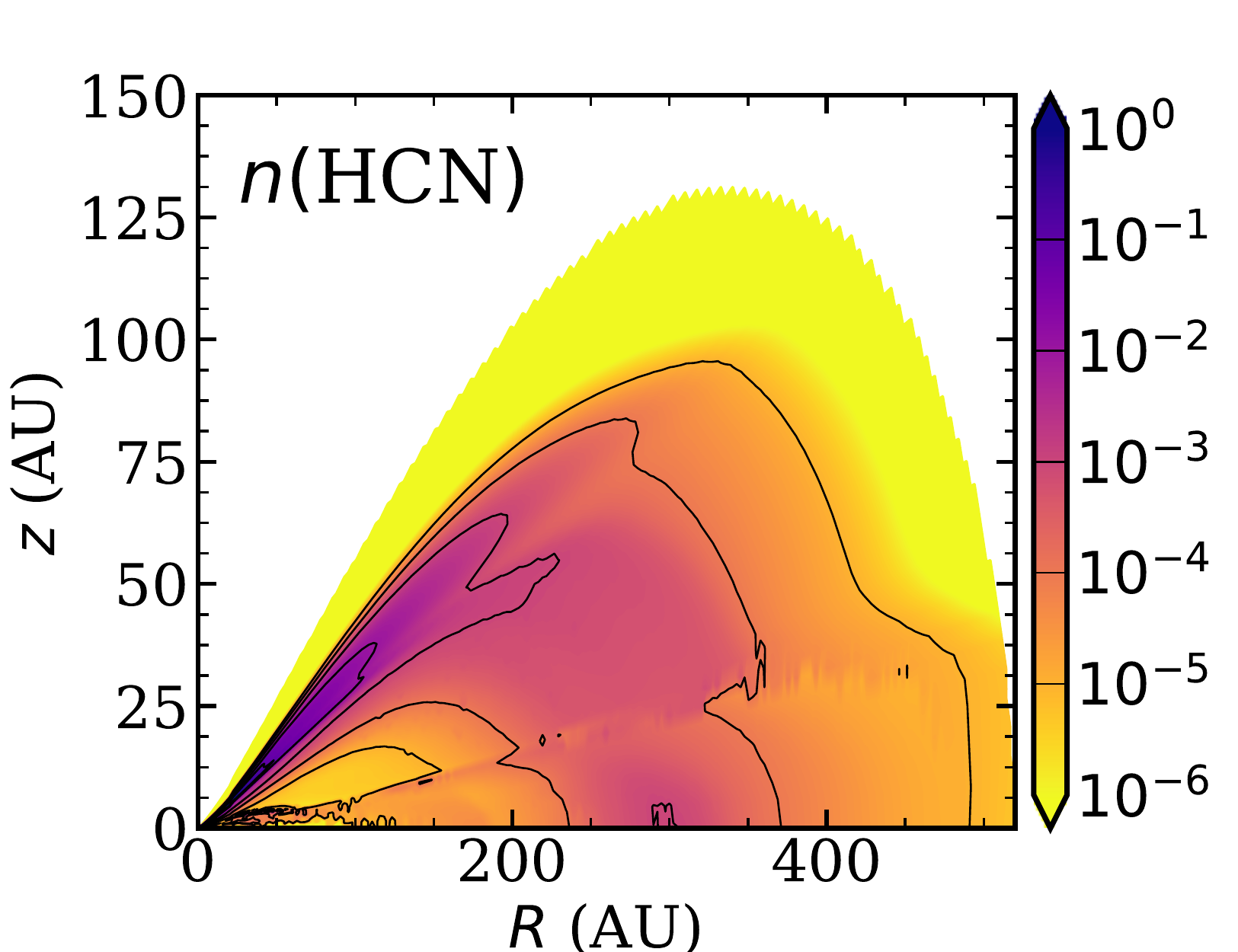}
    \includegraphics[width=0.24\linewidth]{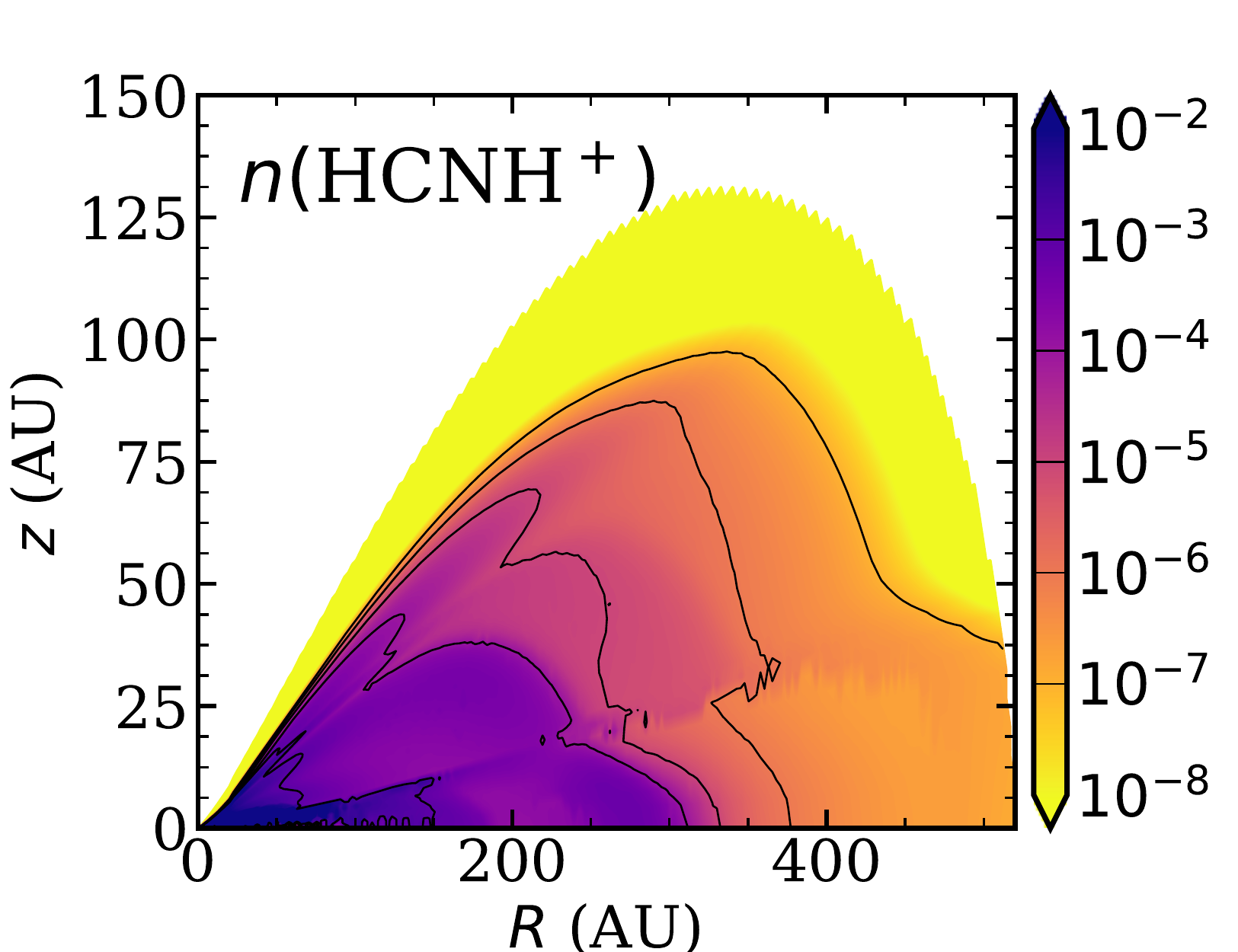}
    \includegraphics[width=0.24\linewidth]{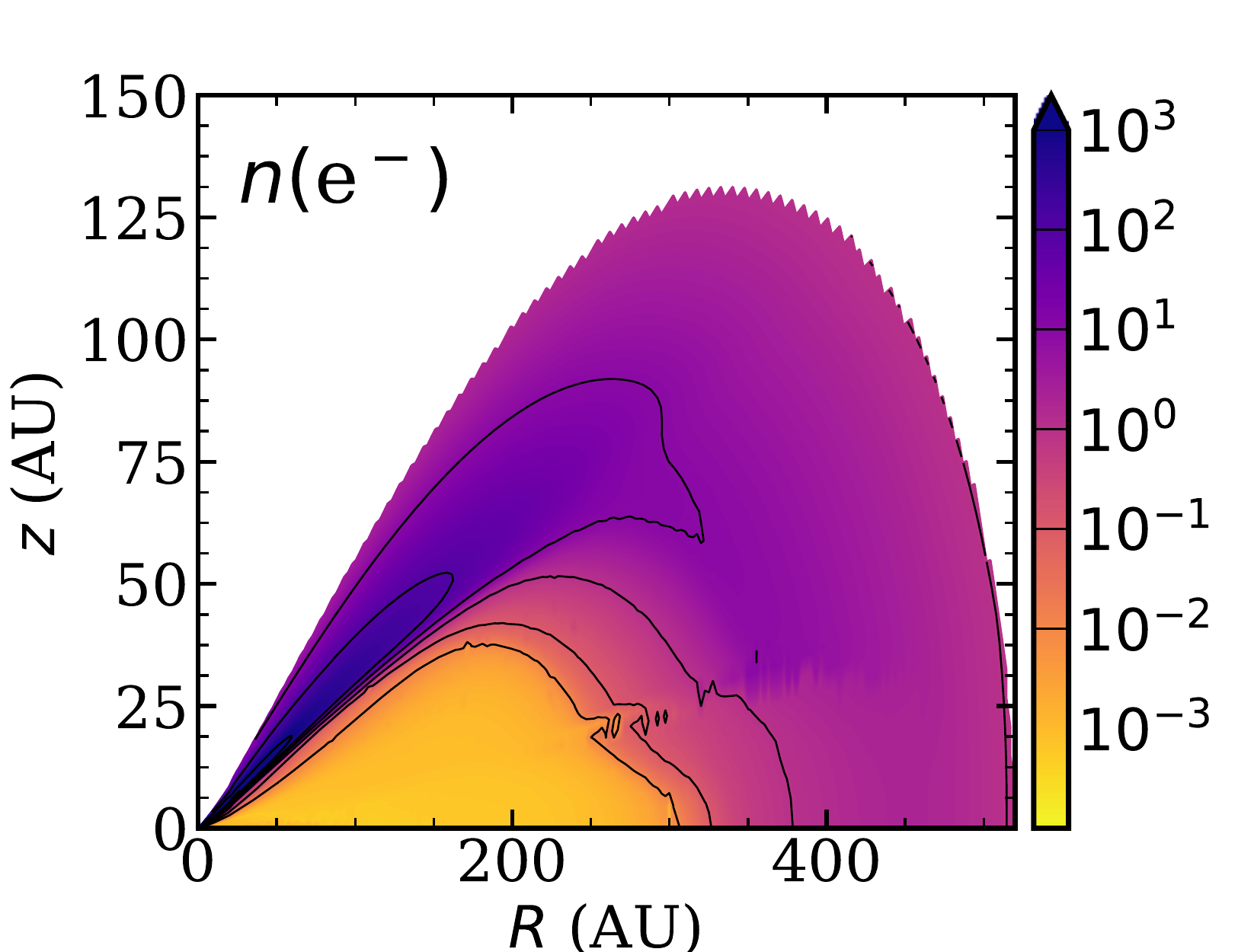}
\caption{\textbf{Upper:} Abundance maps of HNC, HCN, and HCNH$^{+}$ and electrons, the two main reactants for HNC and HCN formation.  Abundance values are relative to $n_{\rm gas}=2n({\rm H_2})+n(\rm H)$. \textbf{Lower:} Gas number density maps of HNC, HCN, HCNH$^{+}$ and electrons. Data are shown for the fiducial model of a $10^{-2}\,\rm M_{\sun}$ disk with $\psi$=0.2, $h_c$=0.1 surrounding a T Tauri star. \label{fig:abun}}
\end{figure*}

\begin{figure}[!th]
\centering
    \includegraphics[width=0.9\linewidth]{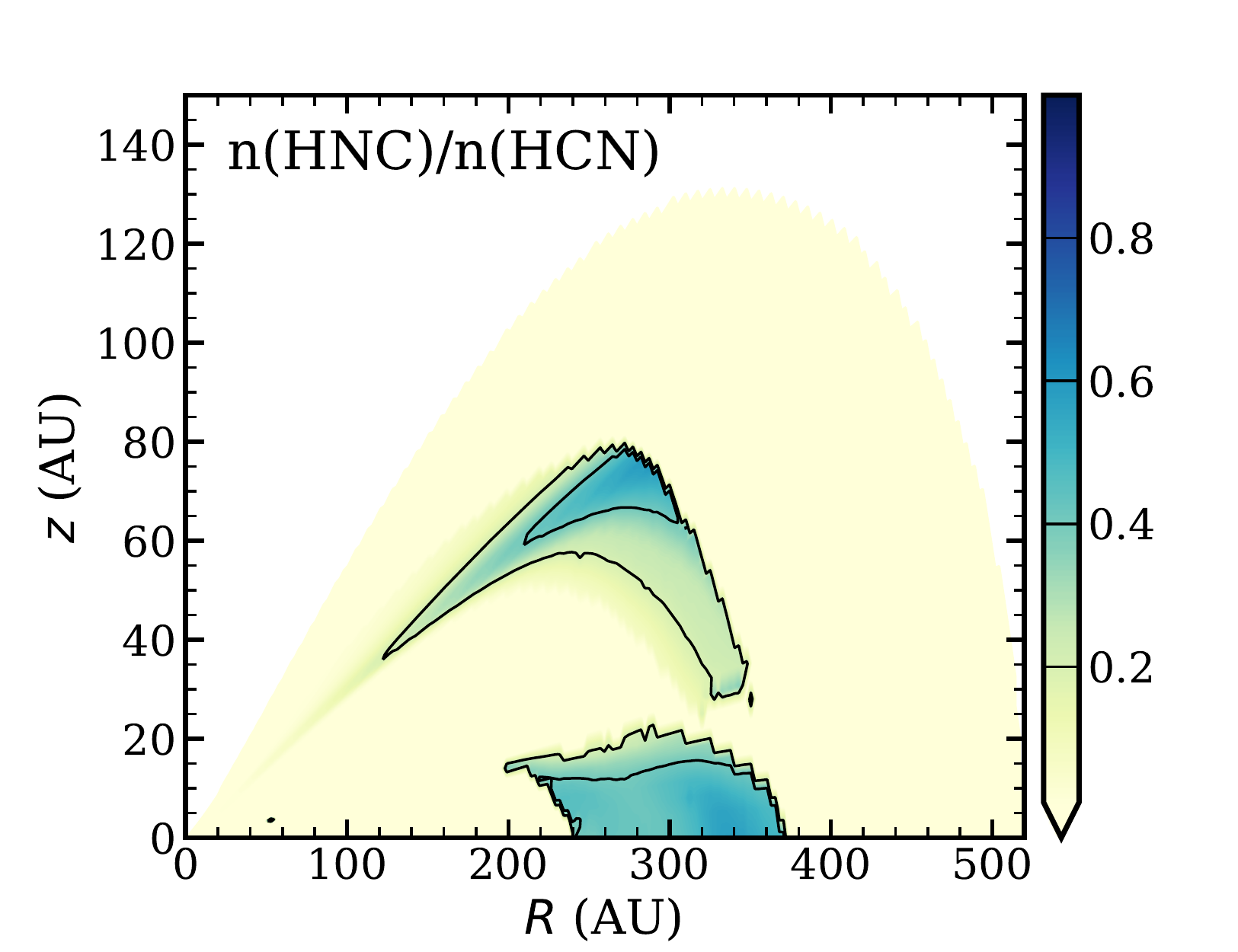} 
\caption{Two-dimensional profile of the HNC-to-HCN abundance ratio in the fiducial model. Only regions with HNC gas number density above $\rm 5\times10^{-5}\,cm^{-3}$ are shown, highlighting the locations where most HNC and HCN emission is produced. \label{fig:2Dratio}}
\end{figure}

\section{Results} \label{sec:results}
In this section, we present the abundance distribution and line emission morphology for HNC and HCN in the fiducial model, which is  a $10^{-2}\,\rm M_{\sun}$ disk with $\psi$=0.2, $ h_c$=0.1 orbiting a T Tauri star. We explore the dependence of line emission on varying disk properties. The 2D distributions of gas density, UV flux, and dust and gas temperature obtained from radiative transfer and thermal balance calculations are shown in Figure~\ref{fig:model} for the fiducial disk model.

\subsection{Abundances} \label{sec:abun}
Figure~\ref{fig:abun} shows the gas densities and abundances (with respect to the total gas density) for HNC and HCN in our fiducial T Tauri disk model. In the hot inner disk ($<$0.5\,au), where CN molecules would be all converted into HCN through reaction with $\rm H_2$ \citep{Walsh2015}, HCN is predicted to be orders of magnitude more abundant than the range covered in Figure~\ref{fig:abun}. We focus on the outer disk where line emission at (sub-)millimeter wavelengths is generated mostly.

Our model suggests similar spatial distributions for HNC and HCN, with regions of higher HNC and HCN abundance and number density located near the disk surface and extending in the midplane from $\sim$200\,au outward (see also Figure~4 in \citealt{Visser2018}). 
Their distributions are expected to closely follow the joint distribution of HCNH$^{+}$ and electrons (see Figure~\ref{fig:abun}) because they are mainly produced through dissociative recombination of HCNH$^+$. At the surface of the disk, HCN and HNC are photodissociated into H and CN, explaining the upper boundaries on the HCN and HNC abundance distributions.

\begin{figure*}[!th]
\centering
    \includegraphics[width=0.42\linewidth, trim=0 0 20 0]{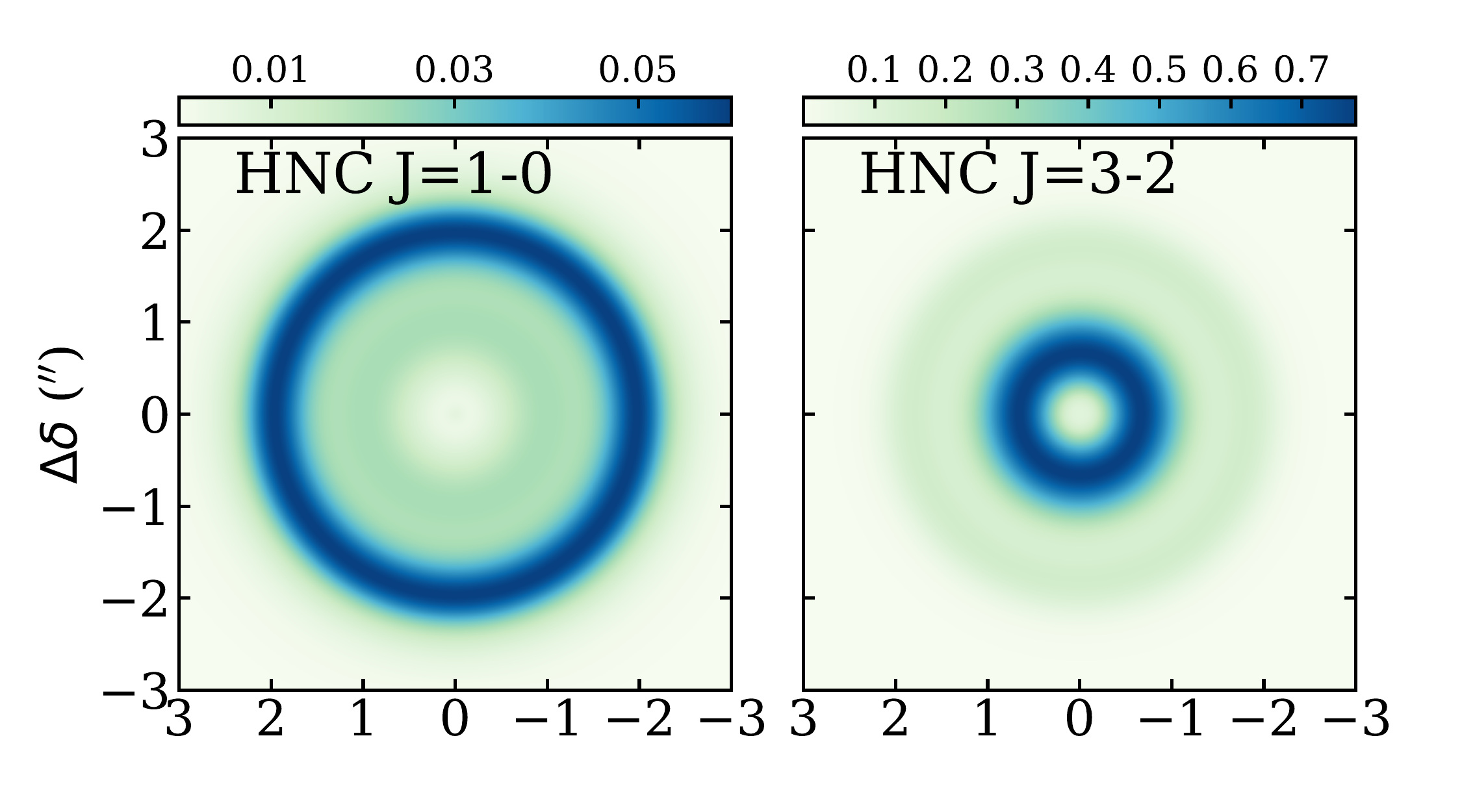}
    \includegraphics[width=0.5\linewidth, trim=0 0 0 0]{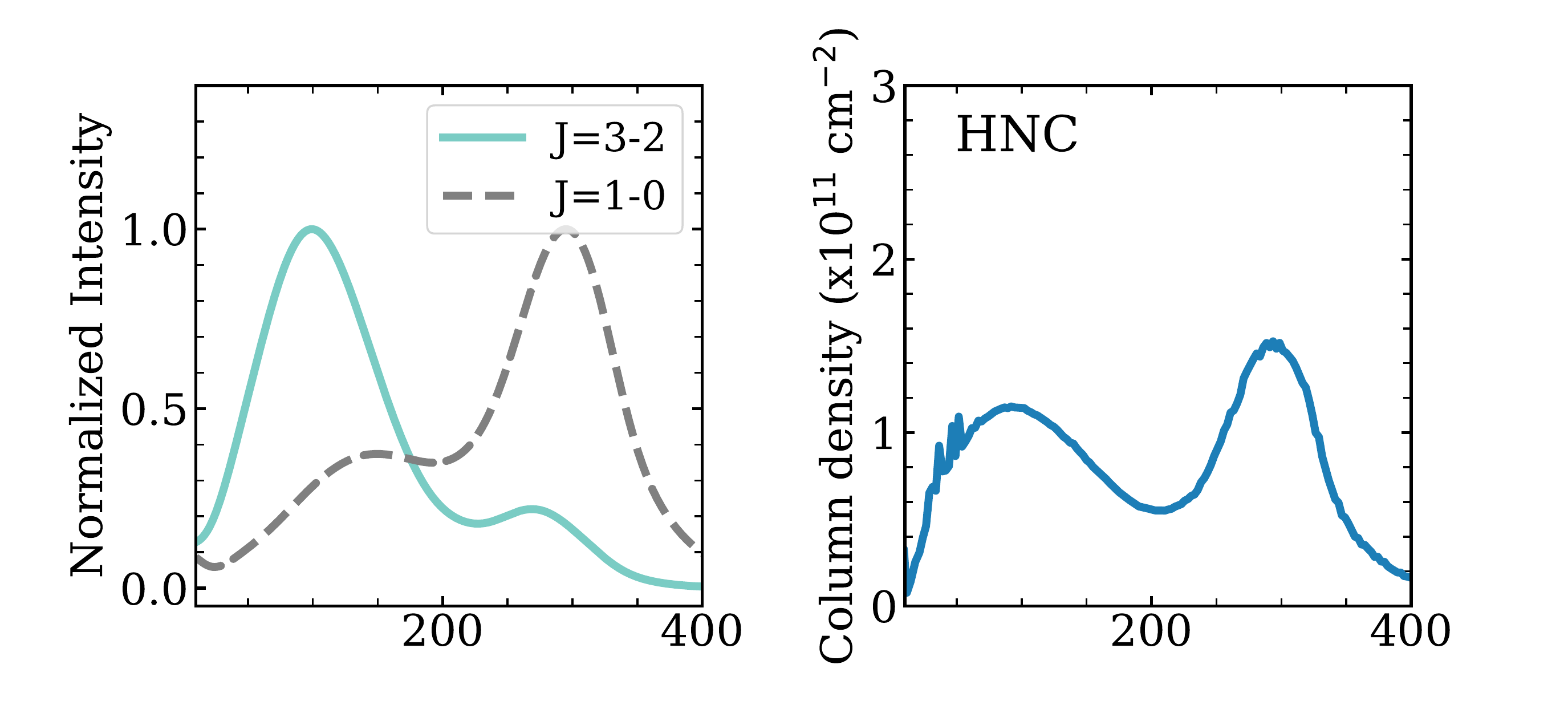} \\ 
    \includegraphics[width=0.42\linewidth, trim=0 0 20 0]{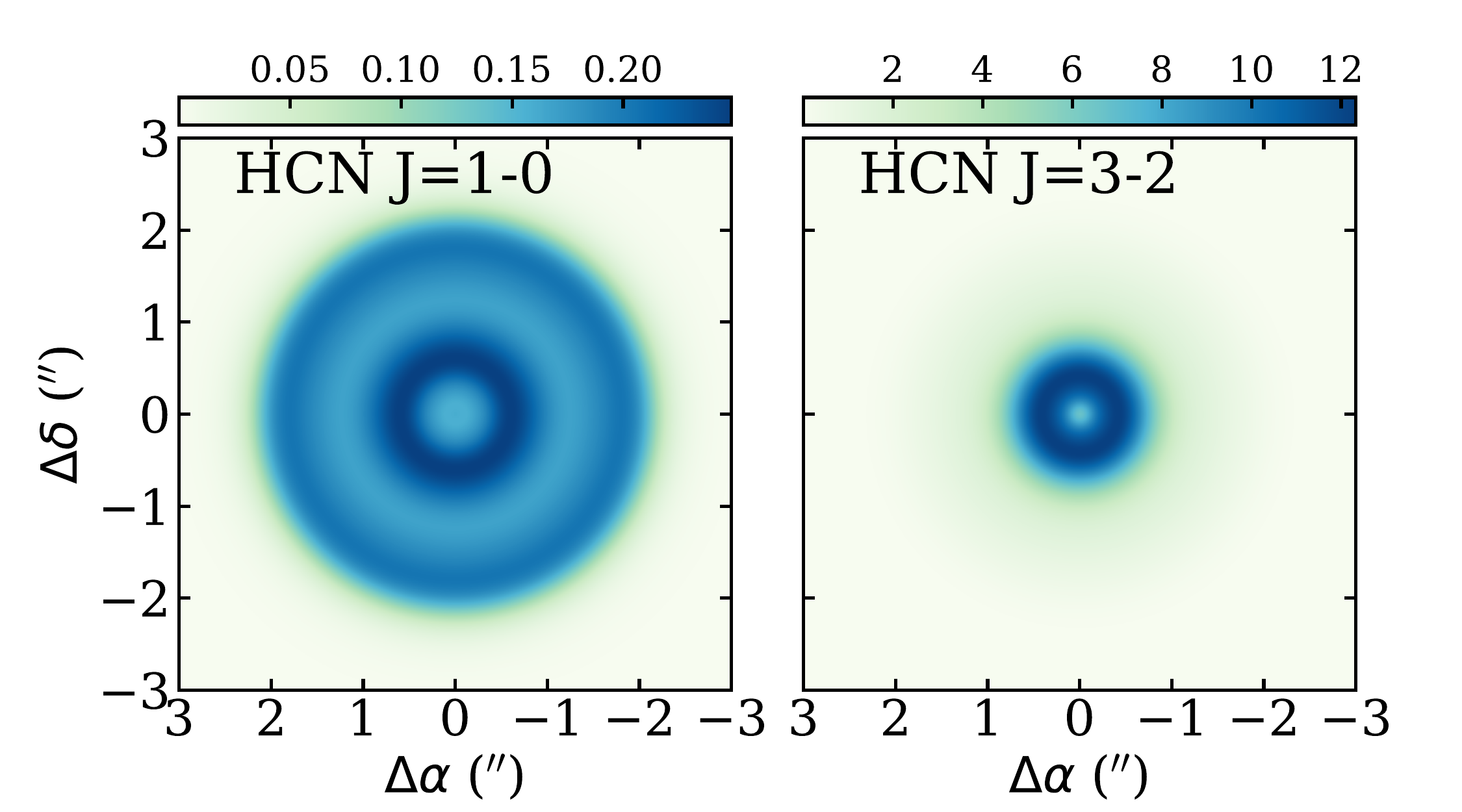}
    \includegraphics[width=0.5\linewidth, trim=0 0 0 0]{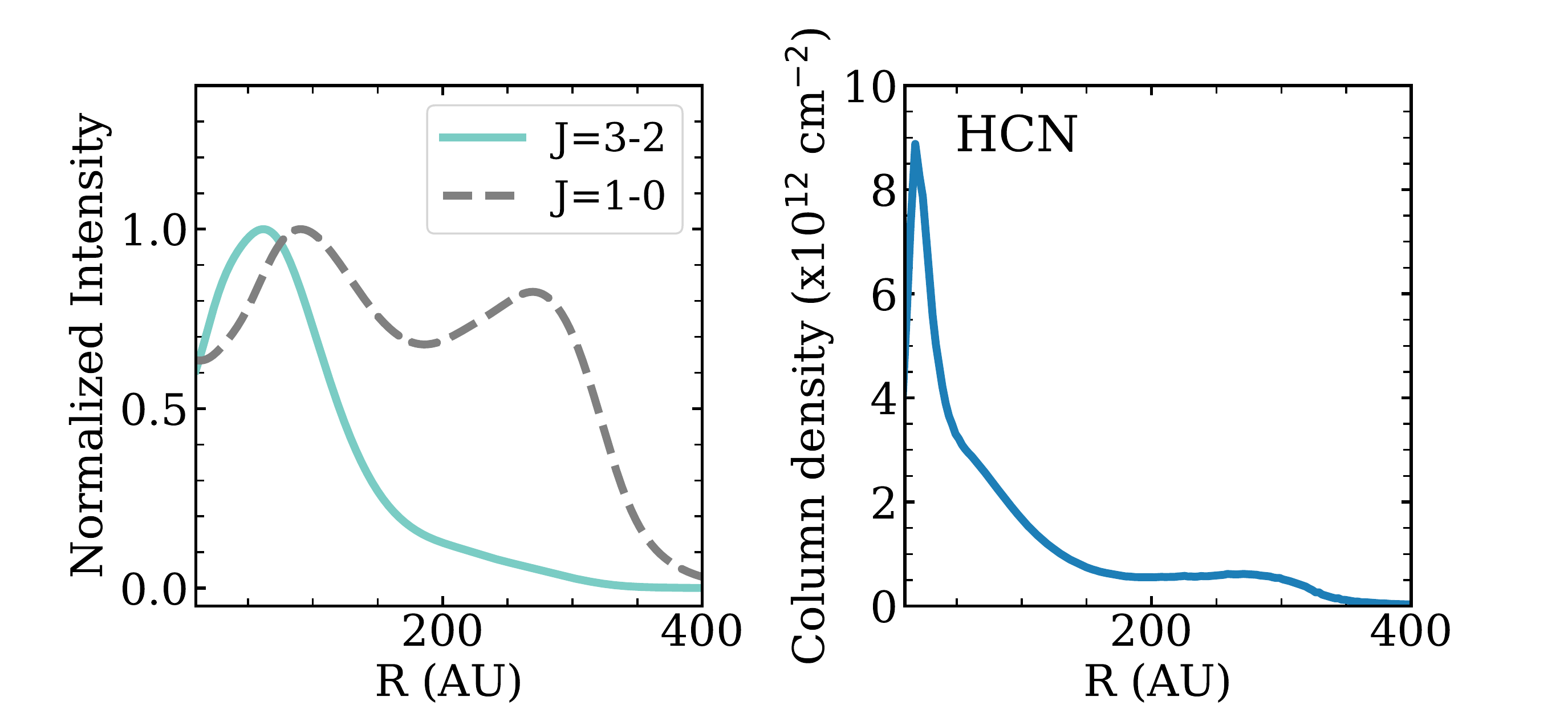} \\ 
\caption{From left to right: Ray-traced images, intensity profile cuts from images, and column density profile. The images are convolved with a 0$\farcs$2 Gaussian beam in units of Jy\,beam$^{-1}$\,km\,s$^{-1}$. Intensity profiles are normalized to the peak of each profile. A distance of 150\,pc has been adopted.  \label{fig:morphology}}
\end{figure*}

Although the HCN and HNC dsitributions are largely similar in our model, there are also important differences.  Figure~\ref{fig:2Dratio} shows the 2D HNC-to-HCN abundance ratio distribution for regions in which the HNC gas number density exceeds $\rm 5\times10^{-5}\,cm^{-3}$. 
Our model predicts overall low abundance ratios ($<0.3-0.4$) across the disk, with localized higher ratios
seen in the cold outer midplane and the warm intermediate layer around 300\,au. In the warm layer, the abundance ratio decreases inward toward the hotter region. This temperature-dependent abundance ratio is expected as the destruction of HNC through reaction~\ref{rx:hnc/h} is more efficient in the warmer regions, leading to lower HNC-to-HCN ratios. The abundance difference between HNC and HCN is also well observed in the column density profiles (see Figure~\ref{fig:morphology}), in which the HCN profile is predicted to peak in the inner disk, while the HNC profile shows a double-peak structure.

\begin{figure*}[!th]
\centering
    \includegraphics[width=0.9\linewidth]{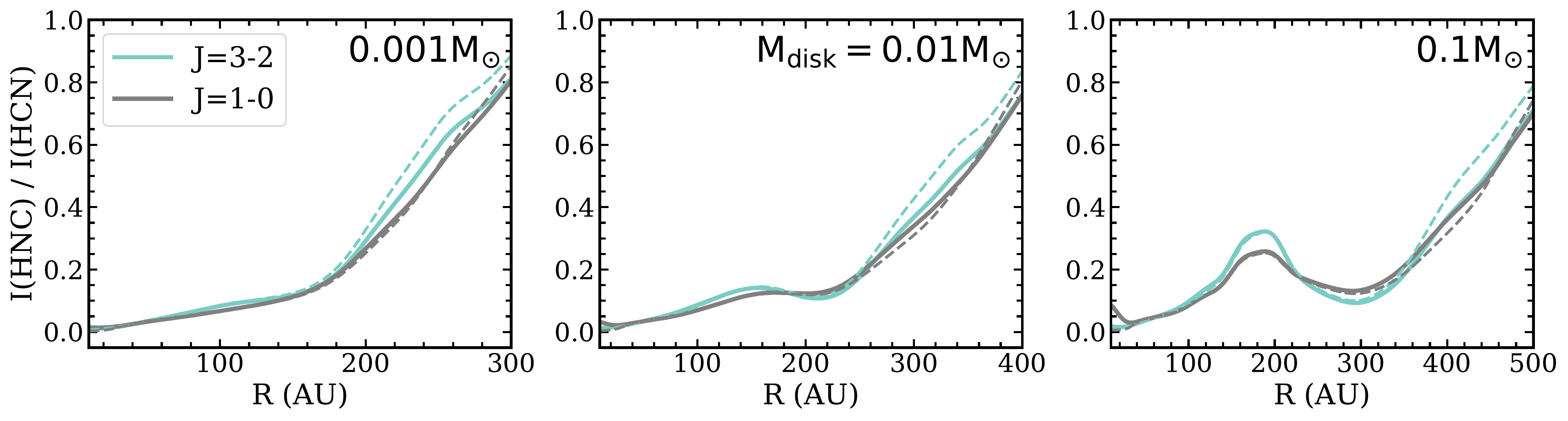}
\caption{Radial profiles of HNC-to-HCN line intensity ratio for the two transitions for the $R_{\rm c}$=60\,au models. The dashed lines are shown for the HNC-to-H$^{13}$CN line ratio, reduced by a factor of 65 (the isotopic ratio), to illustrate the possible optical depth effect. Models with disk masses of 0.001, 0.01 (the fiducial model), and 0.1\,$M_\odot$ are shown from left to right. \label{fig:ratio_radialprofile}}
\end{figure*}

\begin{figure}[!th]
\centering
    \includegraphics[width=0.95\linewidth]{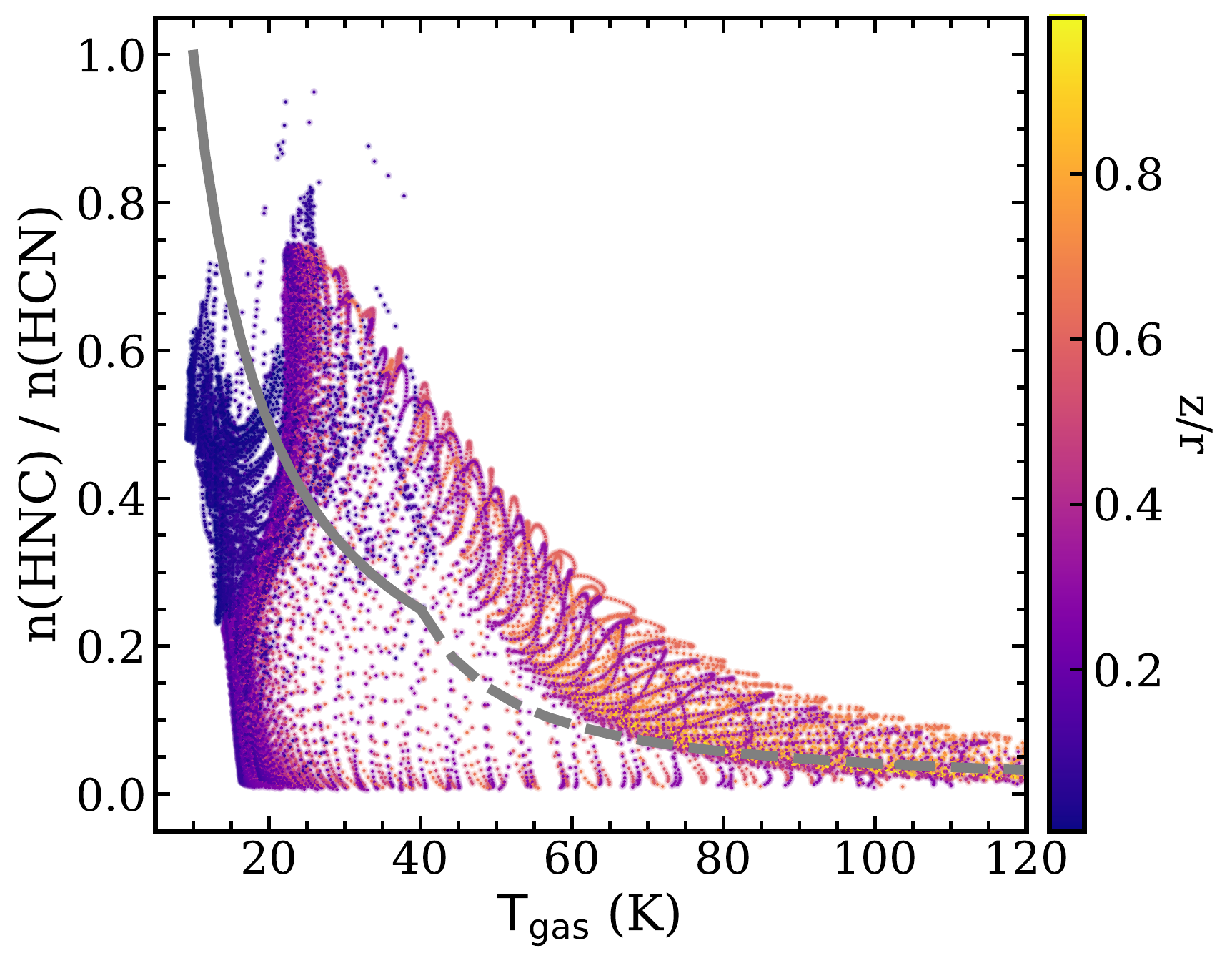}
\caption{Extracted HNC and HCN abundance ratio and the corresponding gas temperature for all models with $\rm M_{gas}=0.01M_{\odot}$, color-coded by scale height $z/r$. Only grids with an HNC gas number density higher than $\rm 1\times10^{-5}\,cm^{-3}$ are selected. The line ratio--gas temperature correlations from \citet{Hacar2020} are overplotted as gray curves. \label{fig:ratio_pixel}}
\end{figure}

\subsection{Emission morphology} \label{sec:morphology}
For a given temperature structure and abundance profile, spectral line emission is obtained through ray-tracing. In this work, we focus on rotational transitions of $J=1-0$ and $J=3-2$ for HNC and HCN, which are close to 90 and 270\,GHz, respectively. These lines are accessible with ALMA at the 1.3 and 3\,mm bands, within which most disk observations are conducted. Figure~\ref{fig:morphology} shows the velocity-integrated intensity maps for HNC and HCN at the two transitions from our fiducial model, convolved with a $0\farcs2$ Gaussian beam. The disk is set to be face-on (inclination of 0$\degr$) and placed at a distance of 150\,pc.

The simulated line emission from both molecules shows a ring-like structure with an emission deficit in the inner disk in the $1-0$ and $3-2$ transitions. Both molecules present a double-ring in both transitions. The relative intensities of the inner and outer rings vary dramatically between species and transitions, however.
For HNC lines, emission peaks at 100\,au for the $3-2$ transition and at $\sim$300\,au for $1-0$ transition; each emission peak corresponds to one of the two peaks in column density profile of HNC. An inner ring is visible for the $1-0$ line and an outer ring for the $3-2$ line, but these are faint compared to the radial emission peaks. For HCN, our model predicts
a prominent ring around 100\,au for both transitions, while a fainter second ring around 300\,au also emerges in the $1-0$ transition. The $3-2$ line presents a small shoulder at 300\,au.

The low-amplitude higher-$J$ emission in the outer disk is likely due to high critical densities of their corresponding transitions (see also the discussion of CN emission in \citealt{Cazzoletti2018}). The critical density of the HNC $3-2$ transition ($\rm \sim8\times10^{6}\,cm^{-3}$ at 20\,K, typical gas temperature around 300\,au) is a few times higher than the local gas density in the outer disk of our fiducial model (see Figure~\ref{fig:model}). We therefore expect the upper level of this transition to be depopulated, resulting in weak line emission (see Figure~\ref{fig:CBF} for the line emission regions). The lower $J$ transitions with lower critical density ($\rm \sim3\times10^{5}\,cm^{-3}$) are still well populated in the low-density outer disk, however, leading to prominent emission there. In addition, the $3-2$ line has a higher upper energy level, which should be substantially less excited in the outer colder disk regions.

The predicted ring-like emission patterns, seen clearly in radial intensity profiles, are consistent with the column density profiles in which peaks of column densities are located off-center (see Figure~\ref{fig:morphology}, right panels).
Ring-like emission for HNC and HCN is naturally produced in our full-disk models (with a centrally peaked surface density profile) because they cannot be abundantly formed in the dense inner midplane region where the electron density is low (see Figure~\ref{fig:abun}). Molecular rings thus do not necessarily correspond to dust rings that are frequently seen in continuum observations. The ring-like structures of HNC and HCN emission seen in the fiducial model are also present in models with smaller characteristic radius ($R_{\rm c}$ = 15, 30\,au), but the ring peak is shifted inward (see Figure~\ref{fig:Rc} for intensity profiles from models with different $R_{\rm c}$). This can be understood when we consider the disk environments that produce the HNC and HCN outer rings: low-density gas where both HCNH$^+$ and electrons are present.

Figure~\ref{fig:ratio_radialprofile} shows the predicted radial profiles of HNC-to-HCN line ratio using the intensity profile cuts from the model images. The HNC-to-HCN line ratio generally increases outward, varying from 0.1 to 0.6 in regions where most line emission originates (50 to 350\,au in the fiducial model). This increasing pattern is consistently seen in our models with different disk masses (Figure~\ref{fig:ratio_radialprofile}). The small bump around 200\,au in the 0.1\,M$_\odot$ disk model might arise because disks with higher disk masses are in general cooler, which slows the HNC destruction down.

The radial profiles of the HNC-to-HCN line ratio from our models are consistent with a scenario in which the line ratio is regulated by temperature in disks because the disk temperature decreases outward. To explore this scenario, we investigated the relation between disk temperature and the HNC-to-HCN ratio more directly. Figure~\ref{fig:ratio_pixel} shows the extracted HNC-to-HCN abundance ratio and gas temperature at each position of the disk for all disk models with the mass of 0.01\,M$_{\odot}$. We focus on disk radii outside of 10\,au where most of HNC and HCN millimeter emission originates. About half of the parameter space is not occupied; high HNC-to-HCN abundance ratios are not expected in the warmest disk regions. Furthermore, the highest achieved HNC-to-HCN abundance ratio in each temperature bin is predicted to decrease with increasing gas temperature. There is no one-to-one relation between temperature and HNC-to-HCN ratio, however.
Figure~\ref{fig:2Dratio} shows that low ratios ($<0.2$) are expected to appear across the disk, corresponding to a wide range of gas temperatures, and high ratios ($>0.4$) would emerge in the outer disk from regions close to the cold midplane and from the warm layer near the disk surface (note the z/r encoded by the color scheme in Figure~\ref{fig:ratio_pixel}). We discuss the possible origin for this distribution in Section~\ref{sec:diss}.

\begin{figure*}[!th]
\centering
    \includegraphics[width=0.9\linewidth, trim=0 20 0 0]{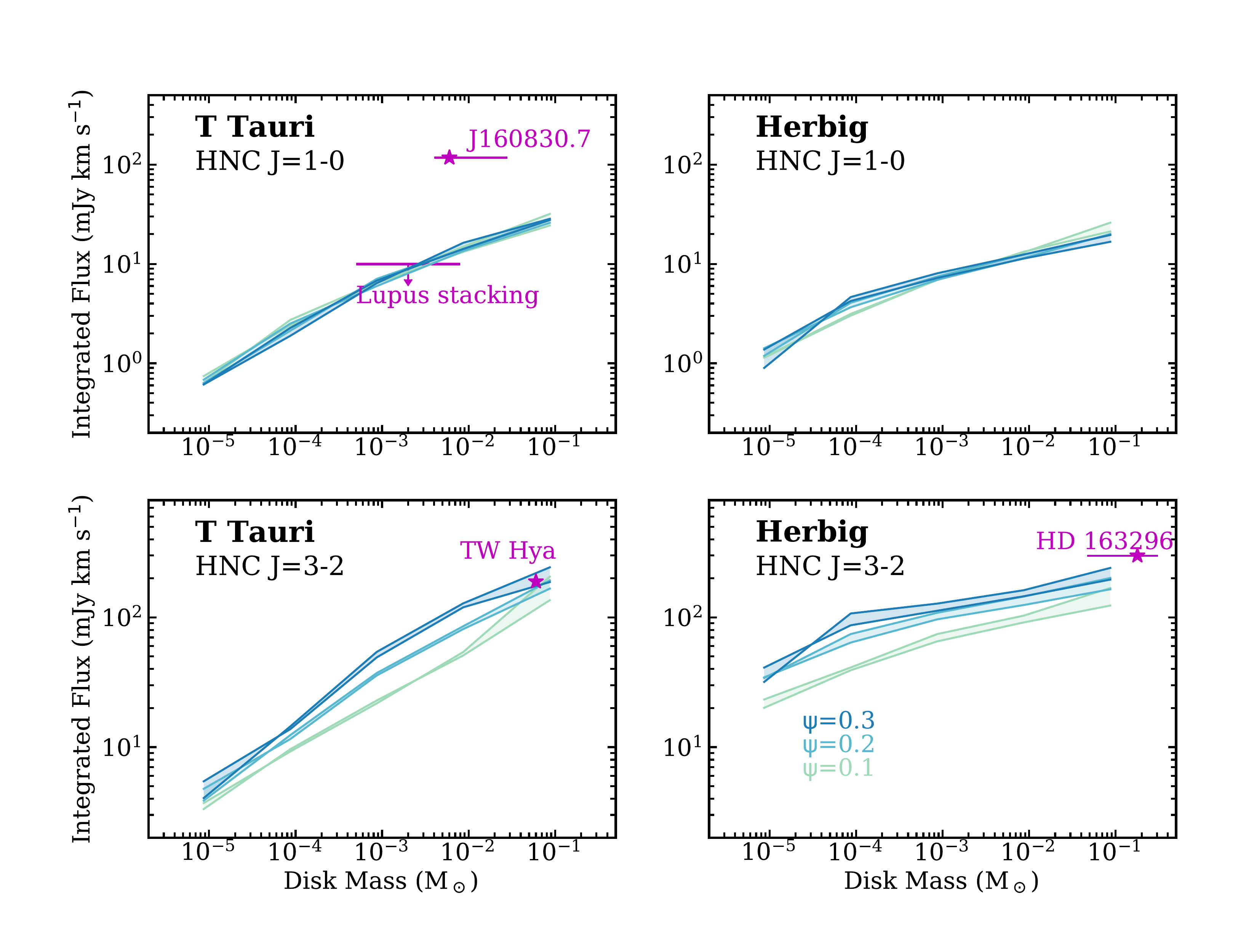}
\caption{Disk-integrated line fluxes (calculated at a distance of 150 pc) of HNC as a function of disk mass (left for disks around T Tauri stars and right for Herbig stars). Colors represent different levels of disk flaring. Upper and lower panels are shown for different transitions. The only detection of HNC $J=1-0$ and the upper limit from stacking analysis for Lupus disks are marked for comparison. Line fluxes for TW Hya and HD 163296 are adopted from \citet{Graninger2015} and are scaled by distances.  \label{fig:flux}}
\end{figure*}

\subsection{Dependence of line fluxes on disk parameters} \label{sec:dependence}
Integrated line fluxes are the most readily obtained observables. Thus it is instructive to explore the dependences of line fluxes on key disk properties. Figure~\ref{fig:flux} shows the predicted general increasing trend of the integrated HNC line fluxes with higher total disk masses. Disks around Herbig stars, with higher UV radiation, are predicted to have brighter (high-$J$) line emission than those around T Tauri stars. The reaction between atomic N and vibrationally excited H$_2$ (via FUV pumping), initializing one major pathway for the HCNH$^+$ formation \citep{Visser2018}, has a strong dependence on UV fluxes, which could then affect the production of HNC and HCN. Our results presented here focus on HNC lines because similar patterns are also expected for HCN lines.

A disk with a larger flaring angle will have more extended disk surface areas exposed to stellar radiation and therefore to UV radiation, which should boost the cyanide chemistry. Figure~\ref{fig:flux} shows that increasing the disk flaring angle results in higher line fluxes of HNC $3-2$. Our model also predicts that the effect of disk flaring (UV radiation) on the emission of HNC $1-0$ line should be negligible, which can be understood when we consider that most of the $1-0$ emission originates closer to the cold disk midplane in the outer disk, which is barely affected by stellar UV radiation (see Figure~\ref{fig:CBF} in the Appendix).

The UV fluxes in our models have contributions from the stellar blackbody spectrum and from an excess UV emission due to accretion. In addition to the default accretion rate of $10^{-8}\,\rm M_{\odot}\,yr^{-1}$, we ran a set of additional models with different accretion rates ($10^{-7}, 10^{-9},$ and $10^{-10}\,\rm M_{\odot}\,yr^{-1}$, with order-of-magnitude differences in FUV fluxes, see \citealt{Visser2018}) to explore the effect of UV fluxes on HNC lines. Similar to what is seen for changing the disk flaring angles, the simulated line fluxes of the HNC $3-2$ transition increase with higher accretion rates, while the fluxes of the HNC $1-0$ lines stay constant (Figure~\ref{fig:acc}). The model-predicted inner ring location for $J=3-2$ line moves outward with increasing UV radiation, consistent with the behavior of CN $N=3-2$ lines as discussed in \citet{Cazzoletti2018}, in which the optimal balance between UV fluxes and gas density is reached farther out when UV radiation is stronger.  This suggests that the disk UV environment could be probed by observations of higher-$J$ HNC and HCN line transitions.

\begin{figure*}[!th]
\centering
    \includegraphics[width=0.9\linewidth]{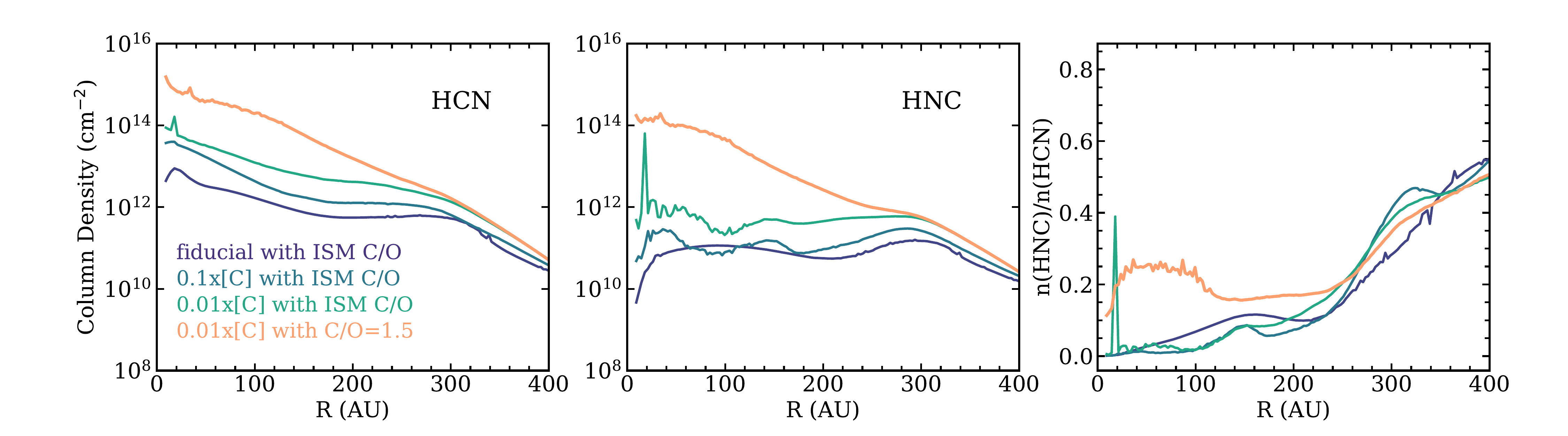}\\
\caption{Column density profiles of HCN and HNC, as well as their ratio profiles, for disk models with different levels of C and O initial abundances: the fiducial model with C/O=0.3 close to the ISM level (in purple), C and O abundances reduced by a factor of 10 (in blue) and 100 (in cyan) from fiducial values, thus with the same ISM C/O, and the C abundance reduced by a factor of 100 from fiducial values with enhanced O depletion to reach a high C/O=1.5 (in orange).\label{fig:cdep_col}}
\end{figure*}

\subsection{Dependence on varying the C and O elemental abundances} \label{sec:C/O}
Recent observations have suggested that substantial 
fractions of the volatile carbon and oxygen are missing in disks, as traced by CO in the submillimeter \citep[e.g.,][]{Favre2013, Miotello2017, Long2017, Zhang2019} and H$_2$O vapor in the FIR \citep{Hogerheijde2011,Du2017}, making the gas-phase carbon and oxygen abundances largely uncertain. 
In addition, in contrast to an ISM-like C/O ($\sim$0.4, similar to the setup in our fiducial model), the bright C$_2$H emission in disks requires further oxygen depletion with a C/O$>$1 \citep{Bergin2016, Miotello2019}. To explore the HNC and HCN line emission under these conditions, we performed three additional carbon- and oxygen-depleted models: (1) The initial overall abundances of carbon and oxygen were depleted by a factor of 10; (2) abundances of carbon and oxygen were depleted by a factor of 100; and (3) under the condition of model (2), oxygen was further depleted to reach a C/O=1.5, all with respect to the fiducial model. The nitrogen abundance was kept the same as in the original model.  

Figure~\ref{fig:cdep_col} presents the column density profiles of HNC and HCN in the depletion models compared to our fiducial model. Our model indicates higher column densities with an increasing level of C and O depletion, especially in the inner disks, and the corresponding line emission is expected to be brighter by a factor of a few. We suspect that the removal of CO modifies the ionization structure in the disk, which then favors the HCN and HNC formation route starting from N$_2$ and He$^+$ (C.12 in \citealt{Visser2018}).  Meanwhile, the destruction of CN with O would slow down with less initial oxygen, leading to a longer CN lifetime and an increased production of HCN and HNC through CN + H$_2$ reaction. The increase in HCN column densities with enhanced C/O as predicted by our models is consistent with the chemical model results from \citet{Cleeves2018}. 
With higher C/O, our model also predicts more centrally peaked column density profile.
The observed diverse morphologies of HCN emission (ring-like or centrally peaked, \citealt{Bergner2019}) might be explained by different C/O in individual disks.

\begin{figure*}[!th]
\centering
    \includegraphics[width=0.4\linewidth]{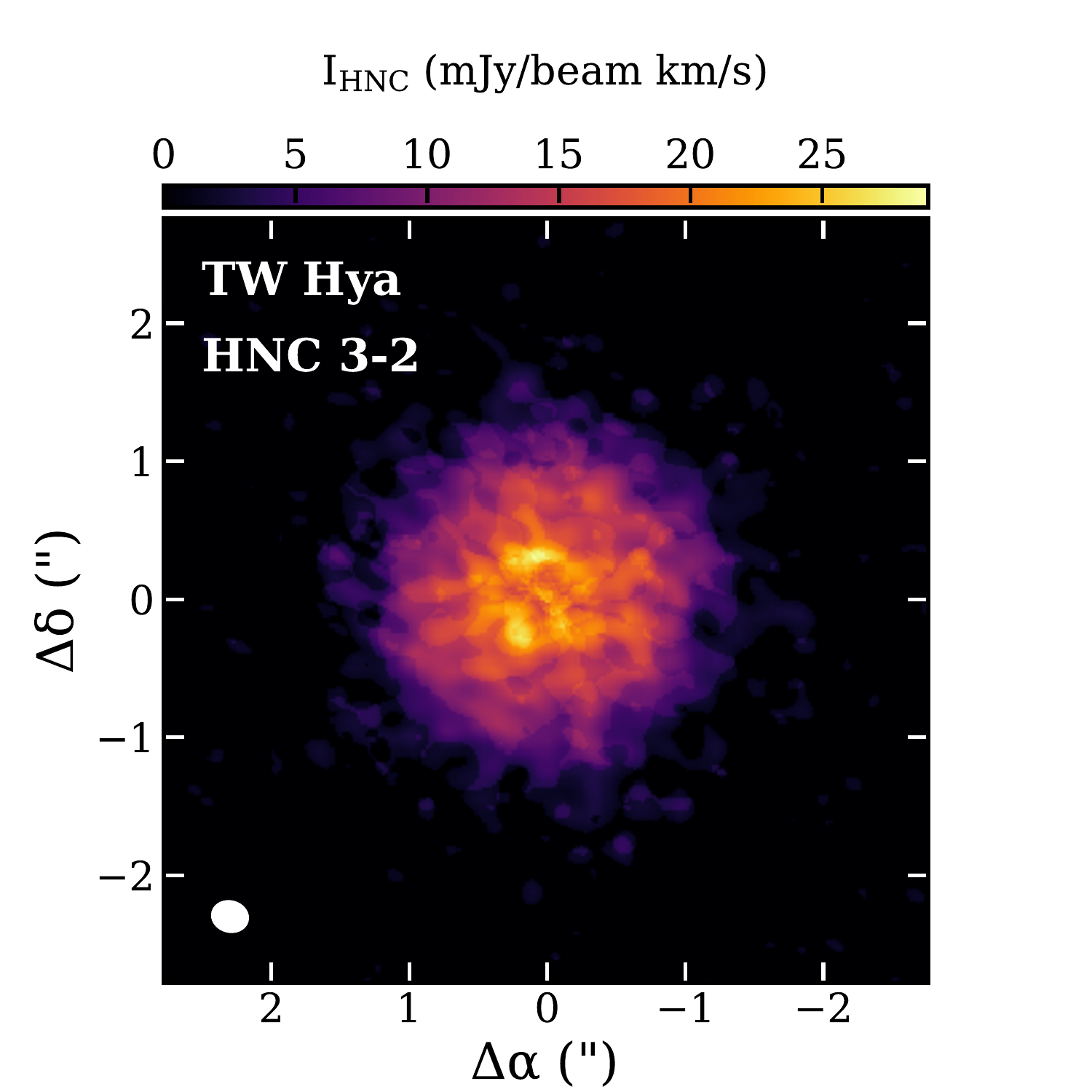}
    \includegraphics[width=0.4\linewidth]{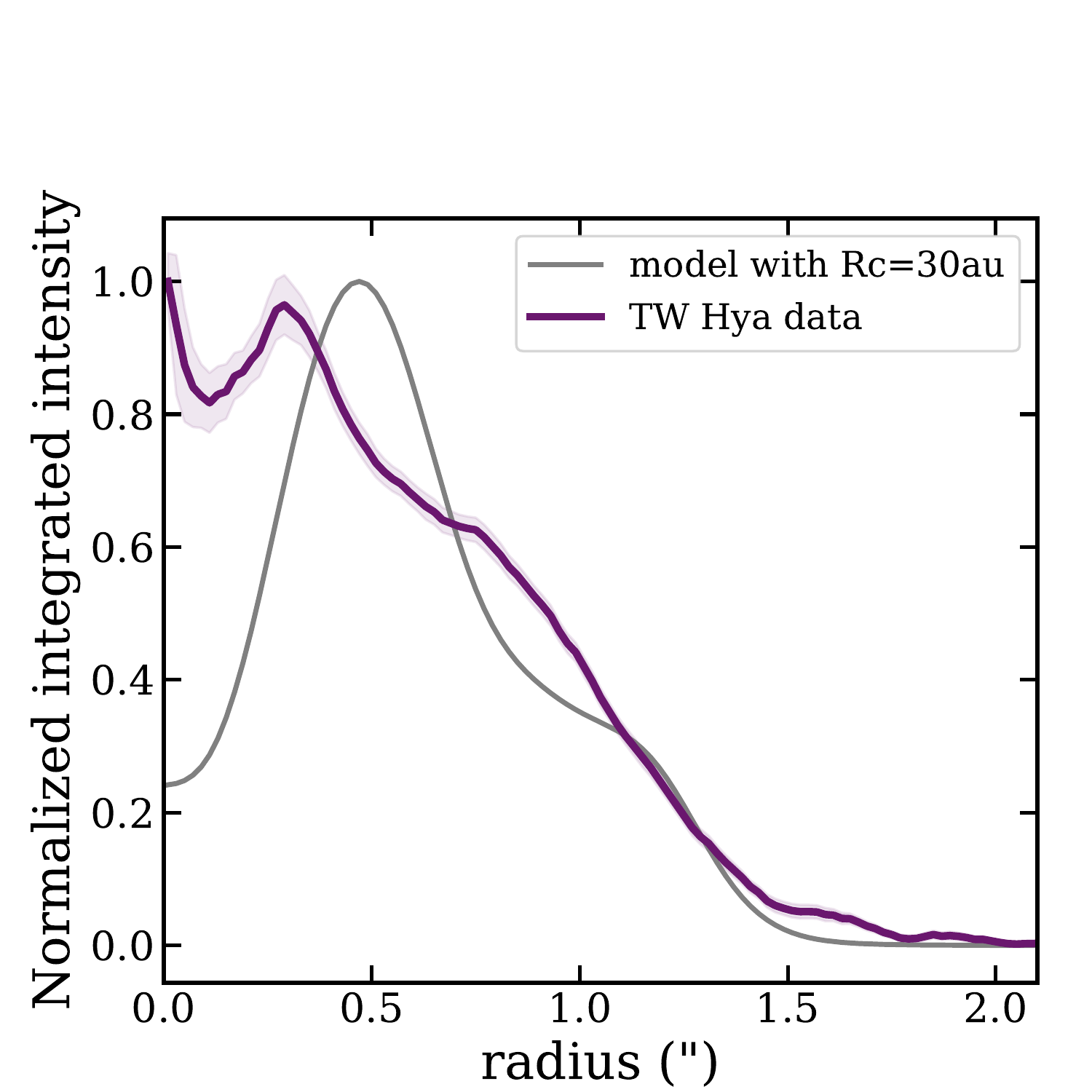} \\
    \includegraphics[width=0.4\linewidth]{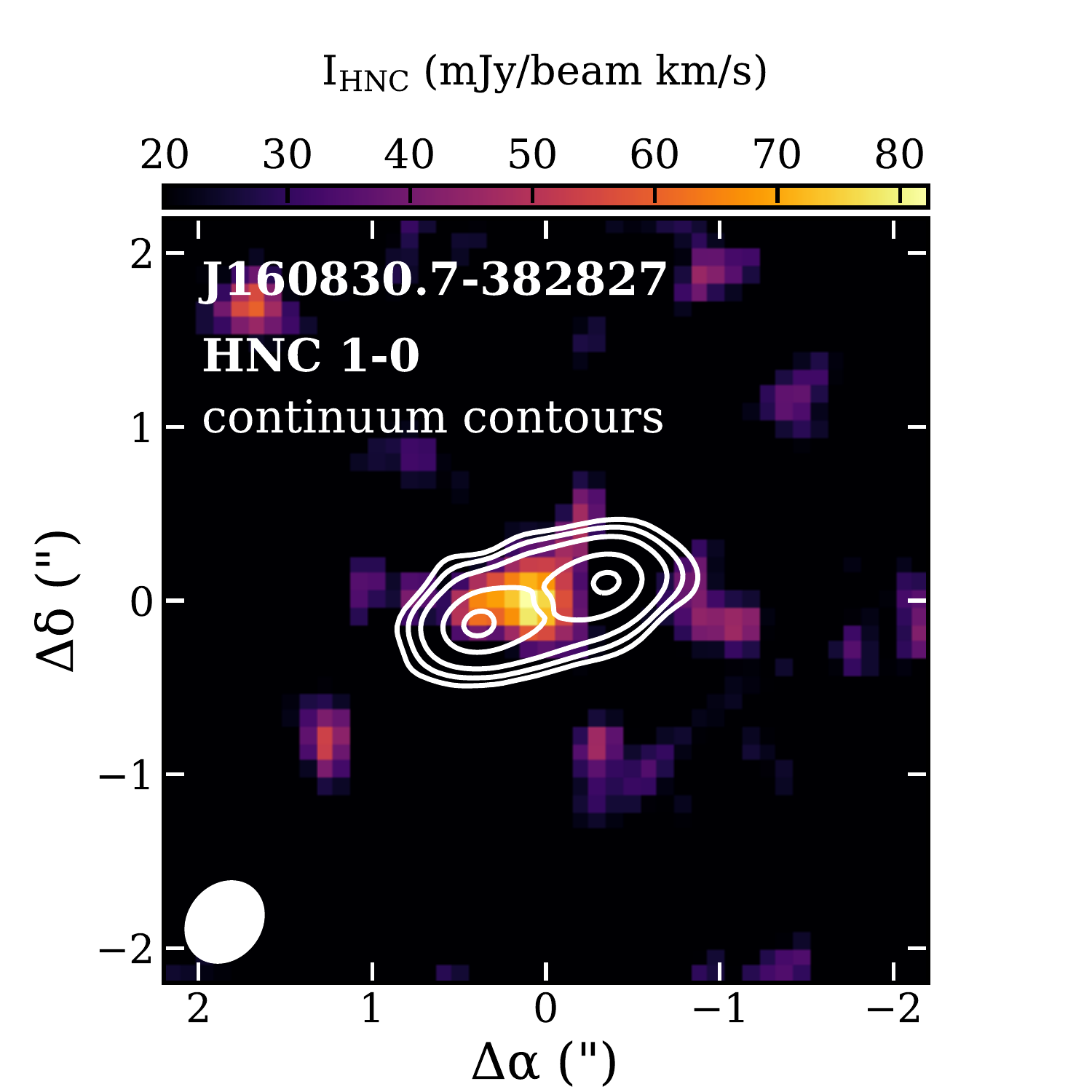}
    \includegraphics[width=0.4\linewidth]{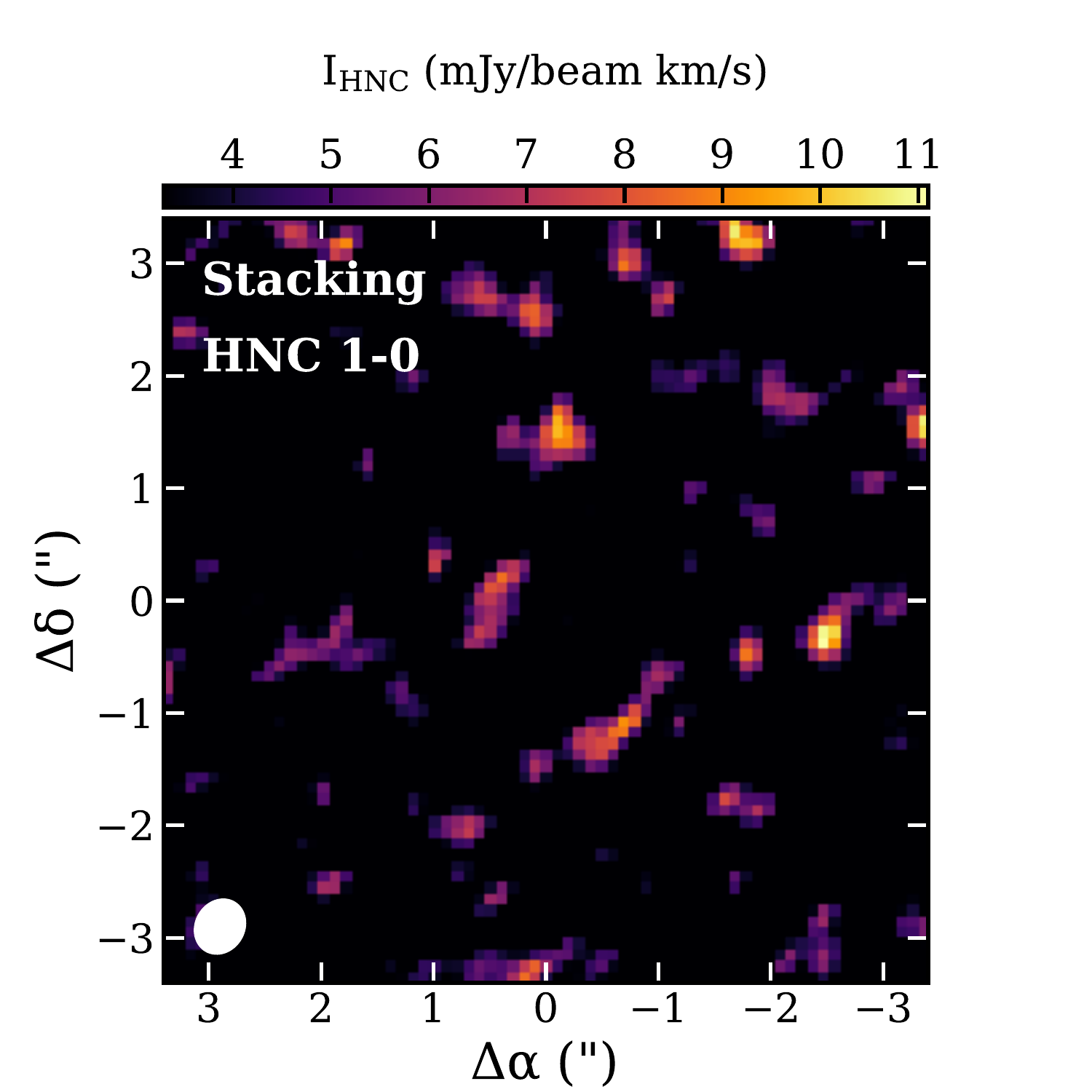} \\
\caption{\textbf{Upper panels:} HNC $3-2$ emission for TW Hya from ALMA observations and the deprojected azimuthally averaged (normalized) radial intensity profile for HNC emission in TW Hya (purple line). A model profile convolved with the same beam size as the ALMA observations is also shown for comparison (gray line). \textbf{Lower panels:} Moment-zero maps for the HNC $1-0$  detection of J160830.7-382827 and stacking results for the remaining 28 disks in the Lupus sample from ALMA observations. White contours depict the continuum emission for the transition disk J160830.7-382827, where dust emission is absent in the disk center, while HNC $1-0$ emission mostly originates within the dust hole. The beam sizes are shown in the left corner of each panel. \label{fig:lupus}}
\end{figure*}

\subsection{Comparisons to observations}

\subsubsection{Comparisons to literature studies}
The brighter HCN lines are commonly targeted and detected in disks \citep{Oberg2010, Oberg2011, Guzman2015, Bergner2019}.
Low- to medium-resolution observations found the disk-averaged column densities for HCN lines to be in the range of (1--10)$\times10^{12}\,\rm cm^{-2}$ for a sample of disks around T Tauri and Herbig Ae stars \citep{Chapillon2012, Bergner2019}. These values are broadly consistent with our models. However, the HCN column densities can reach 10$^{15}\,\rm cm^{-2}$ in the inner 100\,au of disks, as revealed in the recent high-resolution images of the ALMA Large Program (MAPS, V.~Guzman, J.~Bergner, and G.~Cataldi, private communication), which leads to high optical depths of $>$10. High column densities like this could also be obtained from models with C and O depletion (Figure~\ref{fig:cdep_col}). In disk regions in which the observed HCN column density is on the order of 10$^{12}\,\rm cm^{-2}$, the line emission of $J=3-2$ transition is optically thin. Our fiducial model also predicts optically thin line emission of $\tau\sim0.2$ at comparable column densities.

HNC lines are rarely observed and have so far only been detected in a few disks. The first spatially resolved observations of HNC $3-2$ lines were performed toward TW\,Hya and HD\,163296 with the SMA \citep{Graninger2015}. As shown in Figure~\ref{fig:flux}, the distance-scaled disk-integrated line fluxes are consistent with our model predictions for both disks. We adopt the Gaia distance of 59.5\,pc \citep{Gaia2018} and the HD-based gas disk mass of 0.06\,$\rm M_{\odot}$ \citep{Bergin2013} for TW\,Hya. For HD\,163296, we take a range of gas disk masses (0.048--0.31\,$\rm M_{\odot}$) from CO isotopolog and HD observations \citep{Williams2014, Booth2019, Kama2020} with a distance of 101\,pc \citep{Gaia2018}.  Available SMA and IRAM 30m observations provide the disk-averaged HNC-to-HCN line ratio of 0.1--0.2 for the $3-2$ transition in TW Hya and HD 163296, and an upper limit of 0.3 for the $1-0$ transition in HD 163296 \citep{Graninger2015}. These measurements are consistent with our model predictions (see Figure~\ref{fig:lineratio_diskaverage} in the Appendix). As indicated by Figure~\ref{fig:ratio_radialprofile}, a slightly higher line ratio for $1-0$ transition is expected as it mostly emerges farther out in the colder disk region, where the destruction of HNC is mitigated.
The SMA observations with arcsec resolution also indicate ring-like emission for the HNC $3-2$ lines for both disks \citep{Graninger2015}.

\subsubsection{Comparisons to new HNC observations of TW Hya}
The HNC $J=3-2$ line in the disk of TW Hya was observed with ALMA Band 6 receivers on September 30, 2018, with a baseline range of 15--1996\,m (PI: V. Guzman, 2017.1.0156.S). The raw visibilities were first pipeline-calibrated with the required \texttt{CASA} version 5.1.1 and then continuum-subtracted with \texttt{CASA} task \texttt{uvcontsub}. The HNC line data cube with channel widths of 0.15\,km\,s$^{-1}$ was produced with \texttt{tclean} using Briggs weighting of robust=0.5. This returned a beam size of $0\farcs25\times0\farcs20$ and a 1$\sigma$ noise level of 3.5\,mJy\,beam$^{-1}$ per velocity channel. The moment-zero map of the HNC emission and the deprojected azimuthally averaged radial profile, which are created including only pixels above 3$\sigma$ from the velocity range of 1.9--4\,km\,s$^{-1}$, are shown in Figure~\ref{fig:lupus}. 

The line emission in TW Hya exhibits a central component, a bright ring around 0$\farcs$3, and a weaker bump around 1$\farcs$0. This double-ring morphology with the brighter ring at closer radius is consistent with our model predictions as described in Section~\ref{sec:morphology}. In Figure~\ref{fig:lupus} we overplot the radial profile from a model with a smaller characteristic radius $R_{c}=30$\,au (with an otherwise identical parameter setup as the fiducial model) to guide the comparison. We note that the central emission component is not predicted by our fiducial models. The data presented here lack the short-spacing baselines that might affect the radial profile, thus future observations are needed to confirm the presence of this central component. However, TW Hya is known to host an inner disk cavity within 1\,au \citep{Andrews2016}. Its bright C$_{2}$H emission also suggests a strong UV field and a high C/O in the disk \citep{Bergin2016}. A detailed model optimized for TW Hya is required to provide an improved match in emission components, ring locations, and structure amplitudes. This is beyond the scope of this work, however.

\subsubsection{Constraints from new HNC observations of Lupus disks}
The ALMA Lupus disk survey at Band\,3 (PI: M. Tazzari, 2016.1.00571.S) recorded the 3\,mm continuum in frequency-division mode (FDM) with 3840 channels, designed for serendipitous line detections.  The HNC $1-0$ transition was covered in the spectral setup. The observation and data calibration details are described in  \citet{Tazzari2020}. The HNC line data cubes are created with \texttt{tclean} using natural weighting, resulting in a typical beam size of $0\farcs48\times0\farcs40$ with a velocity resolution of $\sim\rm 3.5\,km\,s^{-1}$. With 2--3 min on-source time, we reach a 1$\sigma$ noise level of 2--3\,mJy\,beam$^{-1}$ per $\rm 3.5\,km\,s^{-1}$ channel.

To attempt to detect the HNC $1-0$ line, we generated moment-zero maps by integrating over 1--8\,$\rm km\,s^{-1}$ channels and estimated the noise level from the signal-free zones. The transition disk J160830.7-382827 \citep{vanderMarel2018} is the brightest HNC emission source, with a peak S/N of 4.5$\sigma$. This is incidentally also the brightest CN emission source in the sample \citep{vanTerwisga2019}. Its HNC $1-0$ flux is one order of magnitude above our model prediction (see Figure~\ref{fig:flux}), where we take a source distance of 155\,pc \citep{Gaia2018} and disk gas masses from CO observations (\citealt{Ansdell2016,Miotello2017}).
In addition, HNC $1-0$ emission is detected in the inner disk just within the dust hole (see Figure~\ref{fig:lupus}), as seen also for some other molecules in transition disks (e.g., \citealt{vanderMarel2014}). This mismatch between model predictions and observations is likely explained by the nature of this disk. The models we present here are not appropriate for transition disks, which would have a very different UV environment than full disks.
Disk chemical models that include a central cavity are needed to explore the origin of the observed morphology.

We created the stacked image for the remaining 28 disks and obtained a disk-integrated 3$\sigma$ upper limit of $\sim$10\,mJy\,km\,s$^{-1}$, which agrees with our models in the typical gas disk mass range in the Lupus sample \citep{Ansdell2016}.
The measured CO gas disk sizes for a sample of Lupus disks span from 70 to $\sim$500\,au and are mostly smaller than 300\,au \citep{Ansdell2018}. The CN observations also reveal that a fraction of Lupus disks are compact, consistent with models with a characteristic radius $R_c$=15\,au \citep{vanTerwisga2019}. The HNC $1-0$ emission could therefore originate from different radii in the sample.

\section{Discussion} \label{sec:diss}

\begin{figure}[!t]
\centering
    \includegraphics[width=0.9\linewidth]{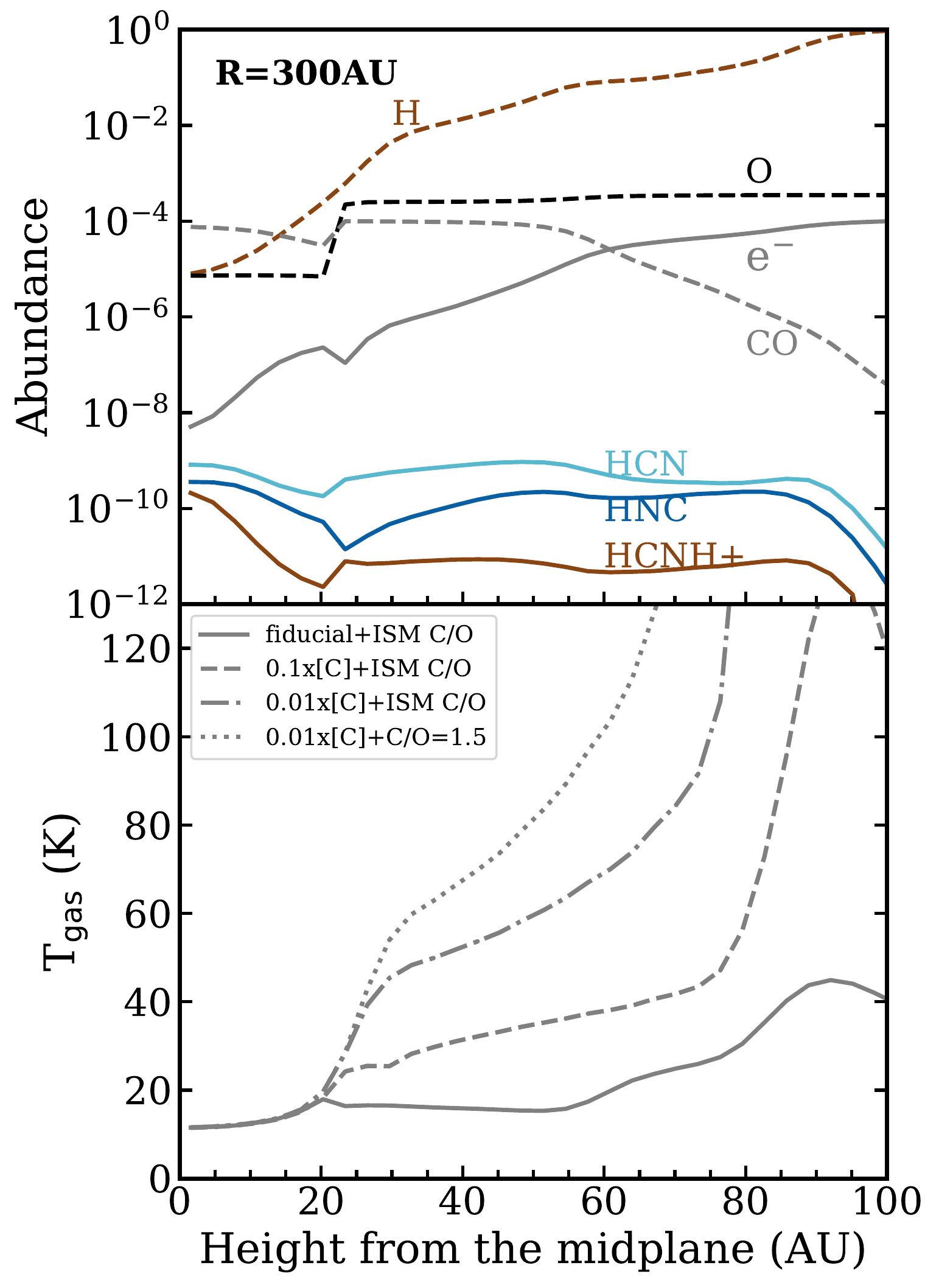}
\caption{\textbf{Top}: Abundance profiles along the vertical disk direction at the radius of 300\,au for HNC and HCN in the fiducial model, as well as species relevant to their formation (HCNH$^{+}$ and e$^{-}$) and destruction (H and O). The drop in CO abundance in the top layer reflects where CO photodissociation becomes important. \textbf{Bottom}: Gas temperature profile along the vertical disk direction at 300\,au for the fiducial model (solid line), as comparisons to carbon and oxygen depletion models (dashed line: a factor of 10; dash-dotted line: a factor of 100; and dotted line: C/O=1.5). \label{fig:vertical_abun}}
\end{figure}

\subsection{HNC-to-HCN intensity ratio as a disk thermometer?} \label{sec:thermometer}
The HNC-to-HCN line ratio has been shown to correlate with temperature in the ISM. The observed line ratio in cold dark clouds is much higher than what is found in the warmer giant molecular clouds \citep{Irvine1984,Goldsmith1986}. A systematic study of the HNC and HCN distribution toward the OMC-1 region found that the HNC-to-HCN ratio is far below 0.1 around Orion-KL, but can reach unity in the colder cloud ridge \citep{Schilke1992}. Recently, \citet{Hacar2020} carried out a dedicated investigation of the HNC-to-HCN ratio in the integral shape filament of Orion and established a strong correlation between the HNC-to-HCN line ratio and the gas kinetic temperature (see the gray curves in Figure~\ref{fig:ratio_pixel}). The HNC-to-HCN ratio has therefore been proposed to have a great potential as a thermometer in interstellar and circumstellar environments. The predicted increasing trend of HNC-to-HCN line ratio with disk radius suggests that the HNC-to-HCN ratio can be used as probe of disk temperature. However, as discussed in Section~\ref{sec:morphology} and Figure~\ref{fig:ratio_pixel}, a one-to-one correspondence between disk gas temperature and HNC-to-HCN ratio is more challenging to establish.

The HNC-to-HCN abundance ratio map shown in Figure~\ref{fig:2Dratio} presents a vertical double-peak pattern that cannot simply be explained by temperature-dependent HNC destruction because the gas temperature increases monotonically toward the surface along the vertical direction. To identify other possible explanations, we examined the abundance vertical cuts at 300\,au for species involved in HNC/HCN formation and destruction (Figure~\ref{fig:vertical_abun}). The disk is assumed to be cold ($\sim$10--20\,K) up to 20\,au above the disk midplane. In this disk region, the reaction rate for the O destruction pathway should rise with increasing temperature, producing a decreasing HNC-to-HCN ratio when moving upward. Above 20\,au, our model predicts a sharp drop in HNC abundance, followed by a gradual increase upward of 25\,au. The sudden decrease in HNC abundance (and HNC/HCN) could be explained by the steep increase in O atom abundance around 20\,au, while we suspect the following increase in HNC abundace (and HNC/HCN) above a height of 25 au may relate to  photodissociation, which gradually takes over as the main destruction pathway for HCN and HNC. Because similar photodissociation cross sections are assumed in the model for the two molecules, we  expect a HNC-to-HCN ratio that gradually approaches unity in photodissociation-dominated regions.
The CO abundance profile, which starts to decrease above 50\,au due to photodissociation, provides the supporting evidence for this hypothesis because the photodissociation of HNC and HCN should occur deeper in the disk than CO photodissociation.

Because the destruction of HNC largely involves C and O, the HNC-to-HCN ratios would be affected by the volatile C and O abundances. In the depletion models discussed in Section~\ref{sec:C/O}, the consumption of HNC through reactions with C and O is suppressed, and the HNC-to-HCN abundance ratio increases with higher level of elemental depletion in regions close to disk midplane (see Figure~\ref{fig:2Dratio_depletion} in the Appendix, and also the right panel of Figure~\ref{fig:cdep_col}). Meanwhile, high HNC-to-HCN abundance ratios in the intermediate disk layer gradually vanish because the gas temperature increases in the depletion models more rapidly (Figure~\ref{fig:vertical_abun}) and the activation of HNC reaction with H would begin deeper in the lower disk layers.

In summary, although the HNC-to-HCN abundance ratio depends on gas temperature in disks, using this ratio as a thermometer in disks will be difficult because the ratio also strongly depends on other parameters, including UV penetration and initial C and O abundances. However, observed abundance ratios above $\sim$0.4 would very likely probe cold or lukewarm ($<$50~K) disk material. A promising way to isolate higher HNC-to-HCN ratios due to low temperatures rather than photodissociation-dominated chemistry is to observe disks with moderate to high inclinations for which the midplane and elevated emitting region could be separated spatially and spectrally. With such observations, the radially resolved HNC-to-HCN ratios in the disk midplane could be deployed as a thermometer, similar to its use in molecular clouds.

Currently available measurements in TW Hya and HD 163296 return disk-averaged HNC-to-HCN $J=3-2$ line ratios of 0.1--0.2 \citep{Graninger2015}. Emission from these shallow observations is therefore likely dominated by the prominent HNC and HCN $J=3-2$ emission component at smaller disk radii. Deep high-resolution observations are required to map the disk radial and vertical thermal structures. The disk around HD 163296, with an intermediate inclination angle, would be an ideal target for this exploration with spatially resolved HCN and HNC observations.

\subsection{Excitation and column density}
The determination of physical properties of the emitting gas depends on the excitation condition of the lines employed. When the emission lines are thermalized and kinetic temperature is known, molecular column density can be derived from the optically thin lines. 
If the LTE condition is not satisfied, the inference of physical quantities would not be so straightforward and detailed radiative transfer calculation with model assumptions are required \citep{vanZadelhoff2001, Teague2018}. As demonstrated in Section~\ref{sec:morphology}, some of the $3-2$ and $1-0$ emission arises from regions in the outer disk, where the critical densities of HNC and HCN are expected to be higher than the local gas density. We find that the excitation temperature for the $1-0$ line in our model is $\sim50\%$ of the gas temperature. While the $1-0$ line may be slightly subthermal, the $3-2$ line at this emitting region would be substantially subthermal because this transition has an even higher critical density. Analysis with HNC and HCN lines should therefore take the non-LTE excitation effects into account for more precise constraints of the disk physical conditions.

\section{Summary}
We presented chemical models of HNC and HCN in protoplanetary disks. Using the 2D thermochemical code DALI, we explored the dependence of HNC and HCN abundance distribution and line emission at (sub-)millimeter wavelengths on various stellar and disk parameters and the potential usage of HNC-to-HCN line ratio as disk temperature probe. We also presented new ALMA observational data of HNC $J=3-2$ in the TW Hya disks and HNC $J=1-0$ in a sample of Lupus disks. Our main findings are summarized as follows: 

\begin{enumerate} \label{sec:sum}
\item HNC and HCN show ring-shaped emission morphology in our full-disk models because these molecules are predicted to be mostly abundant in the warm surface layer and outer disk midplane regions. The observed ring-like structure of the HNC $3-2$ line in the TW Hya disk agrees with our model prediction. These chemical rings are not related to observed dust rings, but probe physical conditions that are relevant to their formation and destruction. 

\item HNC and HCN share similar emission trends to CN, which is very sensitive to stellar UV flux \citep{Cazzoletti2018}. The $J=3-2$ line emission is found to be brighter in model disks with larger flaring angles and higher levels of disk UV radiation and also in disks around Herbig stars than those around T Tauri stars. The $J=1-0$ line is not significantly affected by UV radiation, however, because it primarily originates from the cold midplane in the outer disk.

\item In the models, the column densities of HNC and HCN increase with higher levels of C and O depletion, and they are predicted to be very sensitive to C/O. With a supersolar C/O=1.5, line emission is brighter and tends to peak at the disk center. Radial cuts of HNC and HCN line emission might enable us to probe C/O in disks. 

\item The line fluxes and line ratios from literature HNC observations and the upper limit from new ALMA observations for Lupus disks are consistent with our model predictions. The transition disk J160830.7-382827 is the only detection of HNC $1-0$ from the Lupus survey with a brighter and more centrally peaked line emission than predicted by a full-disk model. 

\item The simulated HNC-to-HCN line ratio generally increases outward, which is consistent with the preferential destruction of HNC in warmer regions. 
This conclusion is robust even for high C and O depletions and C/O$>$1. The precise relation of the line ratio -- gas temperature depends on other parameters such as C and O abundances and UV penetration.

\end{enumerate}

\begin{acknowledgements}
We thank our referee, Joel Kastner, for his constructive comments and suggestions.
We thank Sean Andrews for providing the continuum image for J160830.7-382827 and Richard Teague and Romane Le Gal for helpful discussions. We are grateful to Charlie Qi for fixing the doppler tracking issue in the SMA archival data. 
F.L. acknowledges support from the Smithsonian Institution as the Submillimeter Array (SMA) Fellow. M.T. acknowledges support from the UK Science and Technology research Council (STFC) consolidated grant ST/S000623/1, and by the European Union’s Horizon 2020 research and innovation programme under the Marie Sklodowska-Curie grant agreement No. 823823 (RISE DUSTBUSTERS project).
This paper makes use of the following ALMA data: 2016.1.00571.S and 2017.1.01056.S. ALMA is a partnership of ESO (representing its member states), NSF (USA), and NINS (Japan), together with NRC (Canada), MOST and ASIAA (Taiwan), and KASI (Republic of Korea), in cooperation with the Republic of Chile. The Joint ALMA Observatory is operated by ESO, AUI/NRAO, and NAOJ. 
\end{acknowledgements}

%
%

\bibliographystyle{aa}
\bibliography{aa}{}

\appendix

\section{Line emission region}

\begin{figure*}[!h]
\centering
    \includegraphics[width=0.45\linewidth]{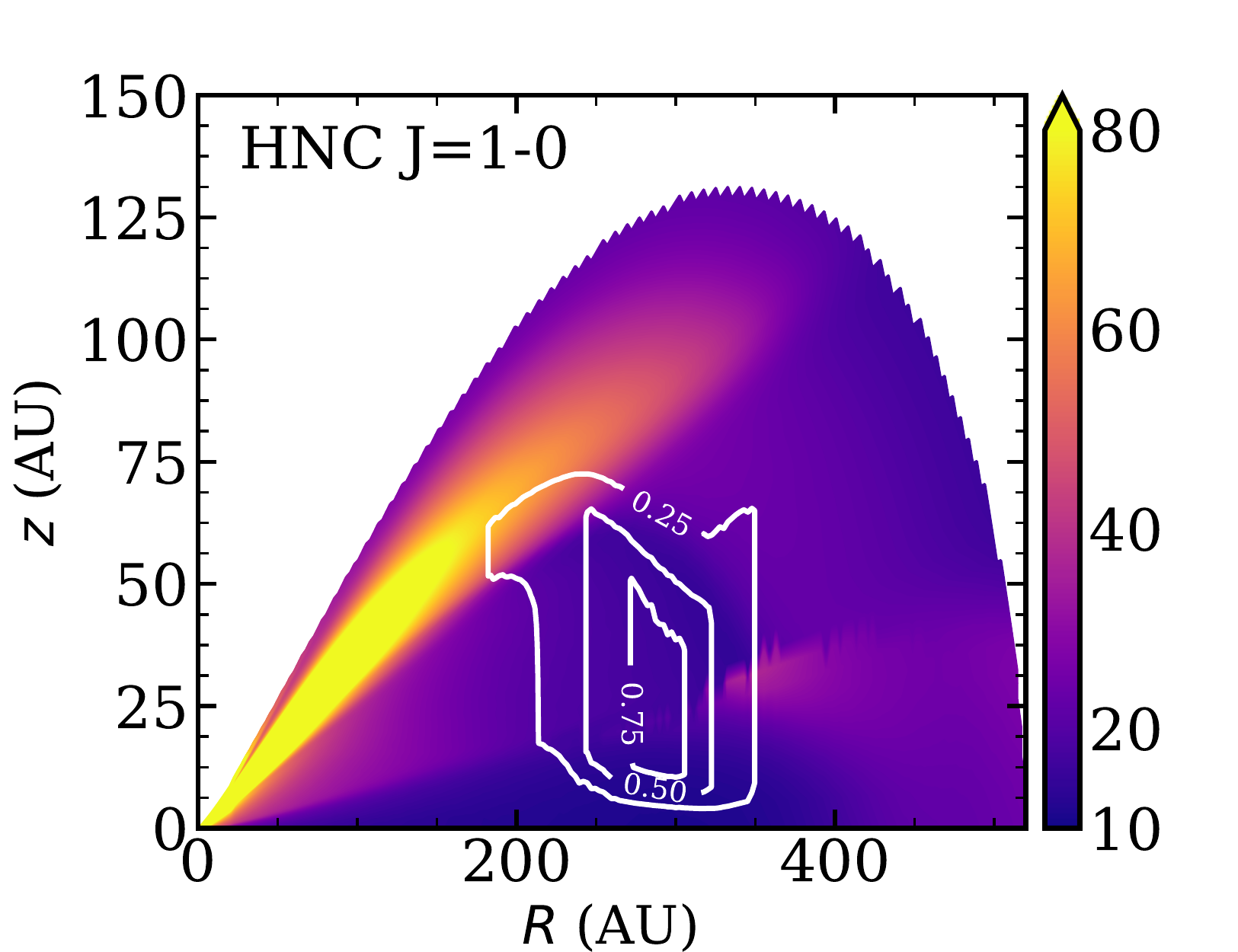}
    \includegraphics[width=0.45\linewidth]{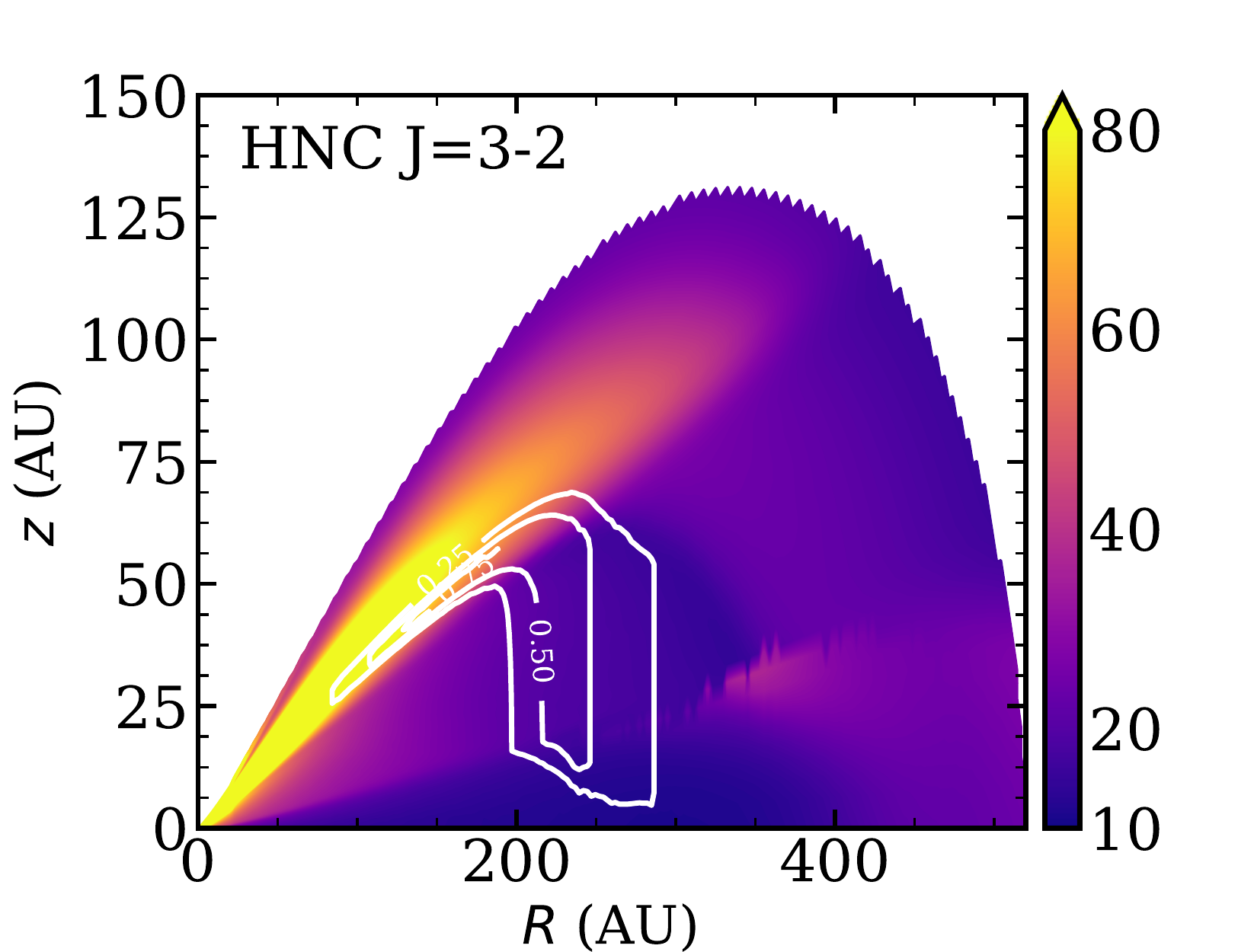}\\
\caption{Contribution function distribution on top of the disk gas temperature distribution. White contours indicate the locations from which 25\%, 50\%, and 75\% of the emission originates. HNC $J=1-0$ mostly emits from the coldest disk region close to disk midplane, while HNC $J=3-2$ largely comes from the warm finger tip. \label{fig:CBF}}
\end{figure*}

A two-dimensional view along the disk radius and the vertical direction of the line emission is shown in Figure~\ref{fig:CBF}, in which the contours indicate the locations from which most of the emission originates. This comparison directly demonstrates the different emission structures of HNC lines seen above: emission of the $J=1-0$ transition predominantly arises from the cold dense region around 300\,au, while emission of the $J=3-2$ transition primarily receives contribution from a warm layer around 100\,au.

\section{Emission dependence on varying disk parameters}
\begin{figure*}[!h]
\centering
    \includegraphics[width=0.9\linewidth]{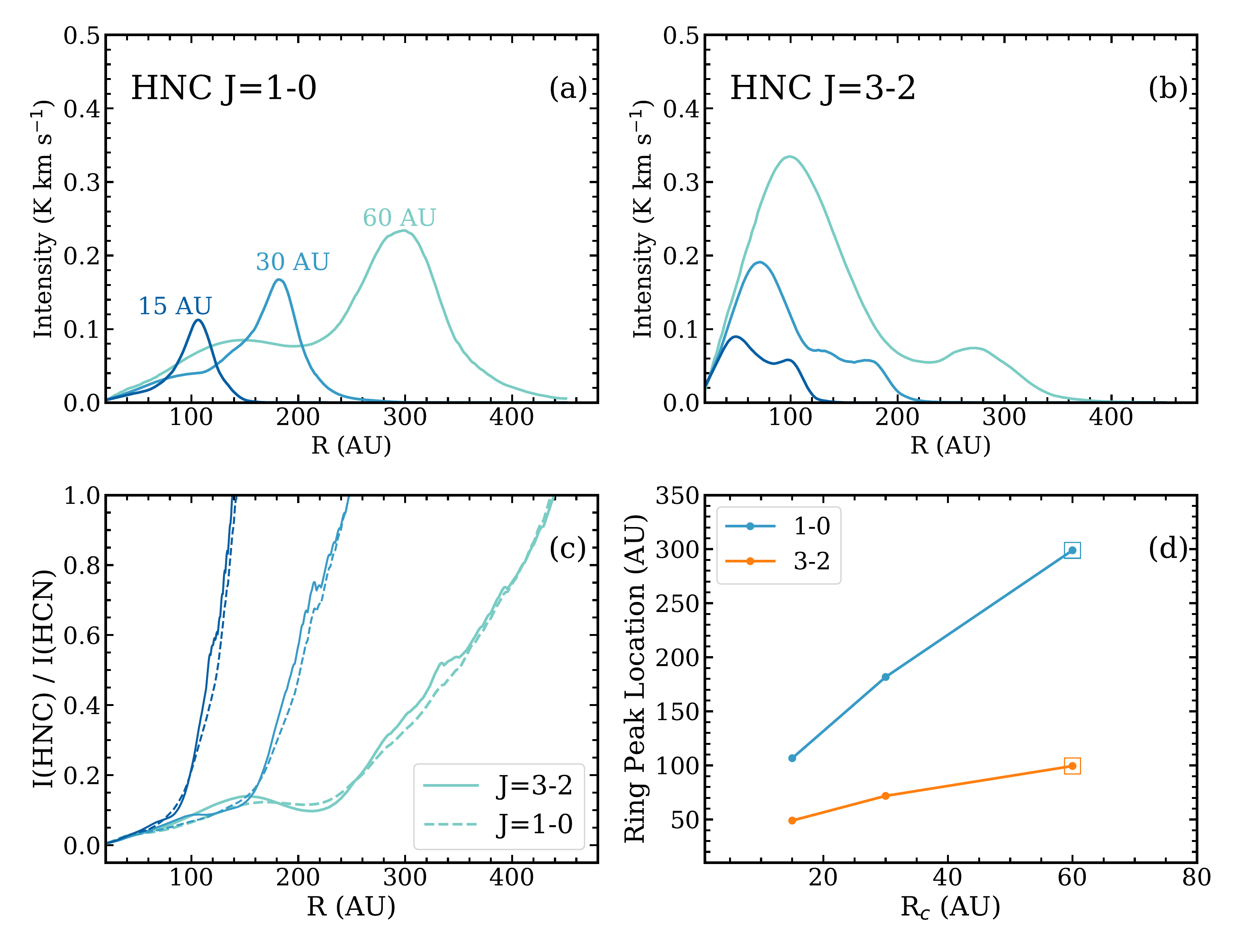}
\caption{Line emission patterns in models with different $R_{c}$ values (disk sizes): (a) Radial intensity profiles for HNC $1-0$ , (b) radial intensity profiles for HNC $3-2$ , (c) HNC-to-HCN intensity ratio profiles, and (d) emission peak location for both transitions. The fiducial model results are marked as open squares.   \label{fig:Rc}}
\end{figure*}

Observations reveal disks with a wide range of radial extensions. We investigated how the HNC line emission varies with different disk sizes, characterized by $R_{c}$ in our models. Based on the fiducial model (e.g., $\psi$=0.2, $h_c$=0.1), we ran additional models with $R_{c}=$15 and 30\,au, keeping the inner disk surface density profile identical (see Figure~B.1 in \citealt{Cazzoletti2018}). Figure~\ref{fig:Rc} shows that the HNC line emission peak shifts inward with decreasing $R_{c}$. This is consistent with the fact that $1-0$ transitions emerge mostly in the outer midplane, tracing the disk size. The shifts of the line peak in the $3-2$ transitions could be understood as showing that the balance of FUV flux and gas density is reached closer in the smaller disks with lower density. The HNC-to-HCN line intensity ratio profiles behave similarly in the region in which most emission originates. 

\begin{figure*}[!h]
\centering
    \includegraphics[width=0.9\linewidth]{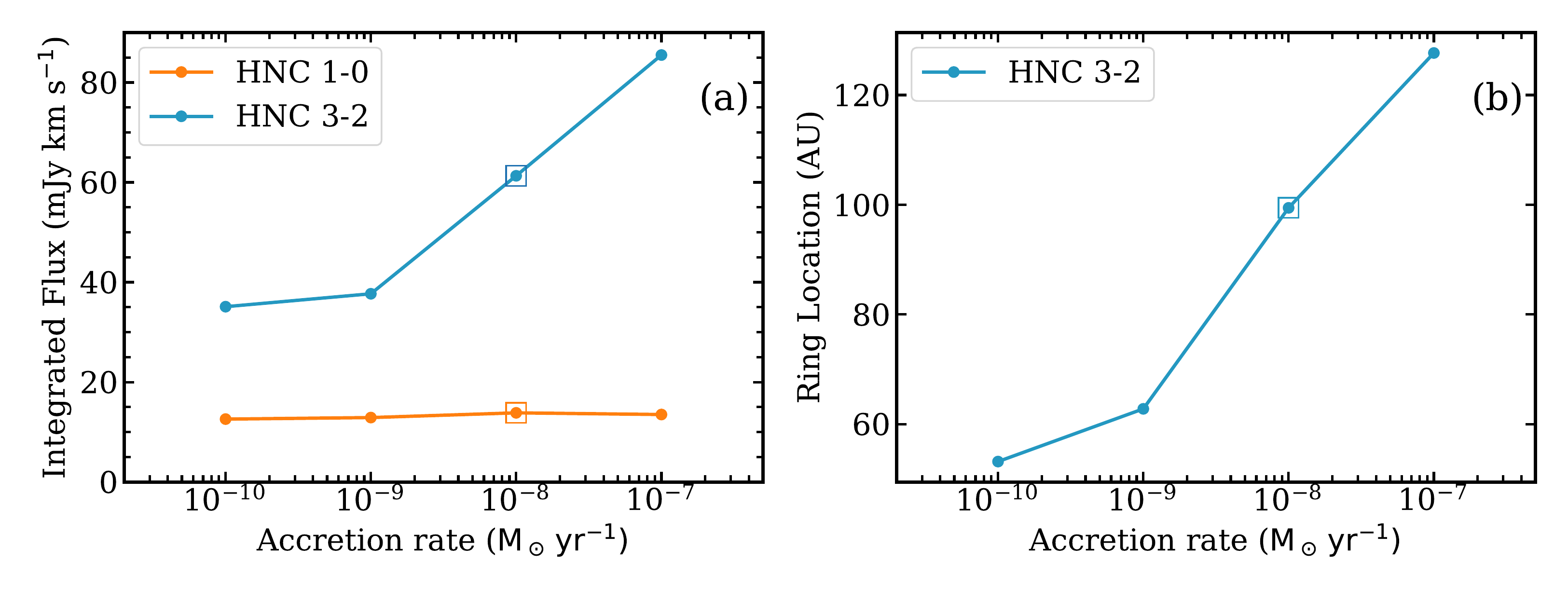}
\caption{(a) Disk-integrated line fluxes for models with different accretion rates, which are a measure of UV luminosity. (b) Emission peak location change for the HNC $3-2$ line, which is sensitive to UV radiation. The fiducial model results are marked as open squares.  \label{fig:acc}}
\end{figure*}

The UV radiation is very important for cyanide chemistry, and stellar accretion could introduce excess UV fluxes on top of the stellar spectrum. As discussed in Section~\ref{sec:dependence} and shown in Figure~\ref{fig:acc}, brighter $3-2$ line emission is observed in systems with a higher accretion rate and higher UV fluxes, with the line peak shifting outward. The $1-0$ lines are barely affected by UV radiation.

\section{HNC-to-HCN line ratio compared to observations}

\begin{figure*}[h]
\centering
    \includegraphics[width=0.43\linewidth]{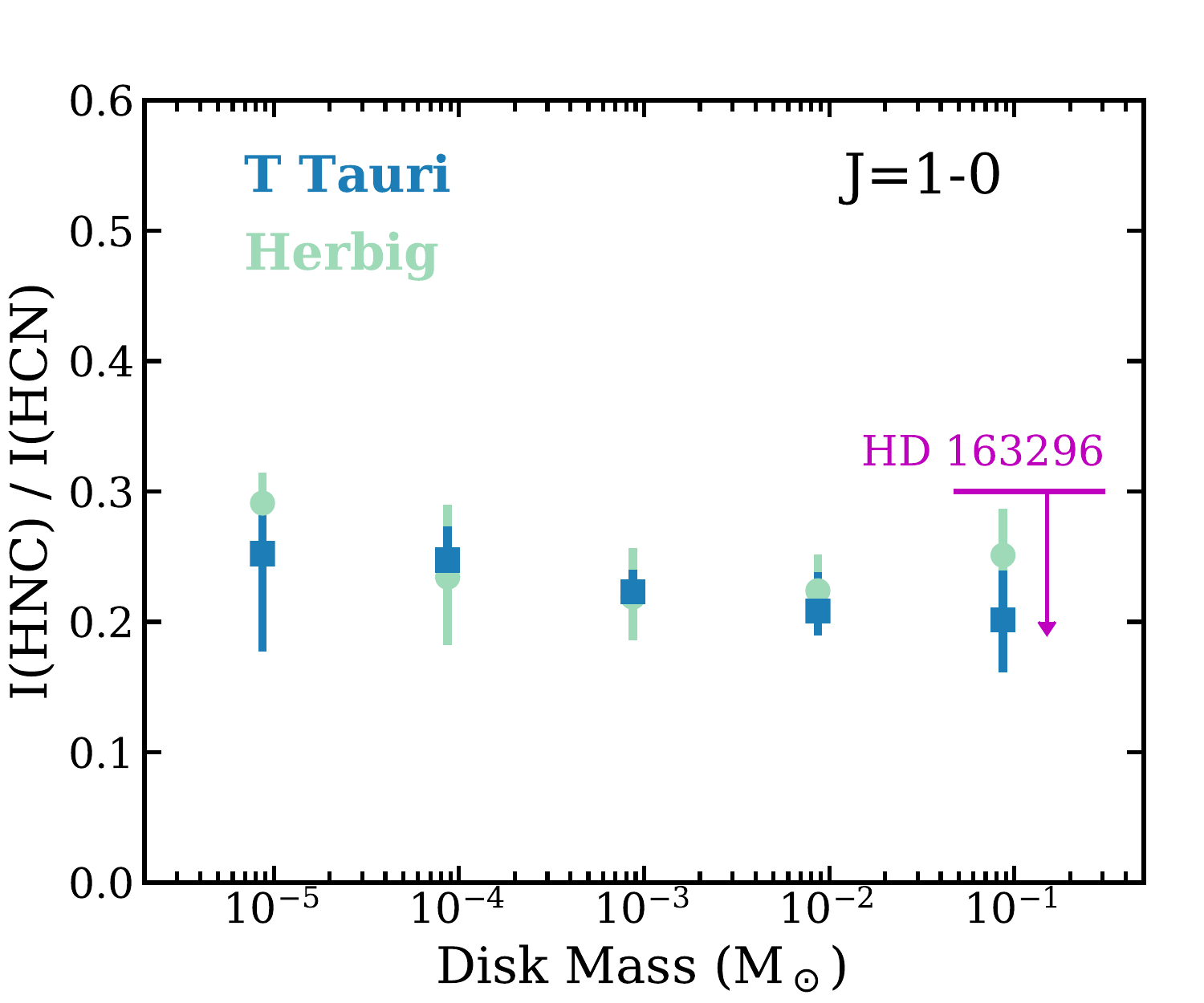}
    \includegraphics[width=0.43\linewidth]{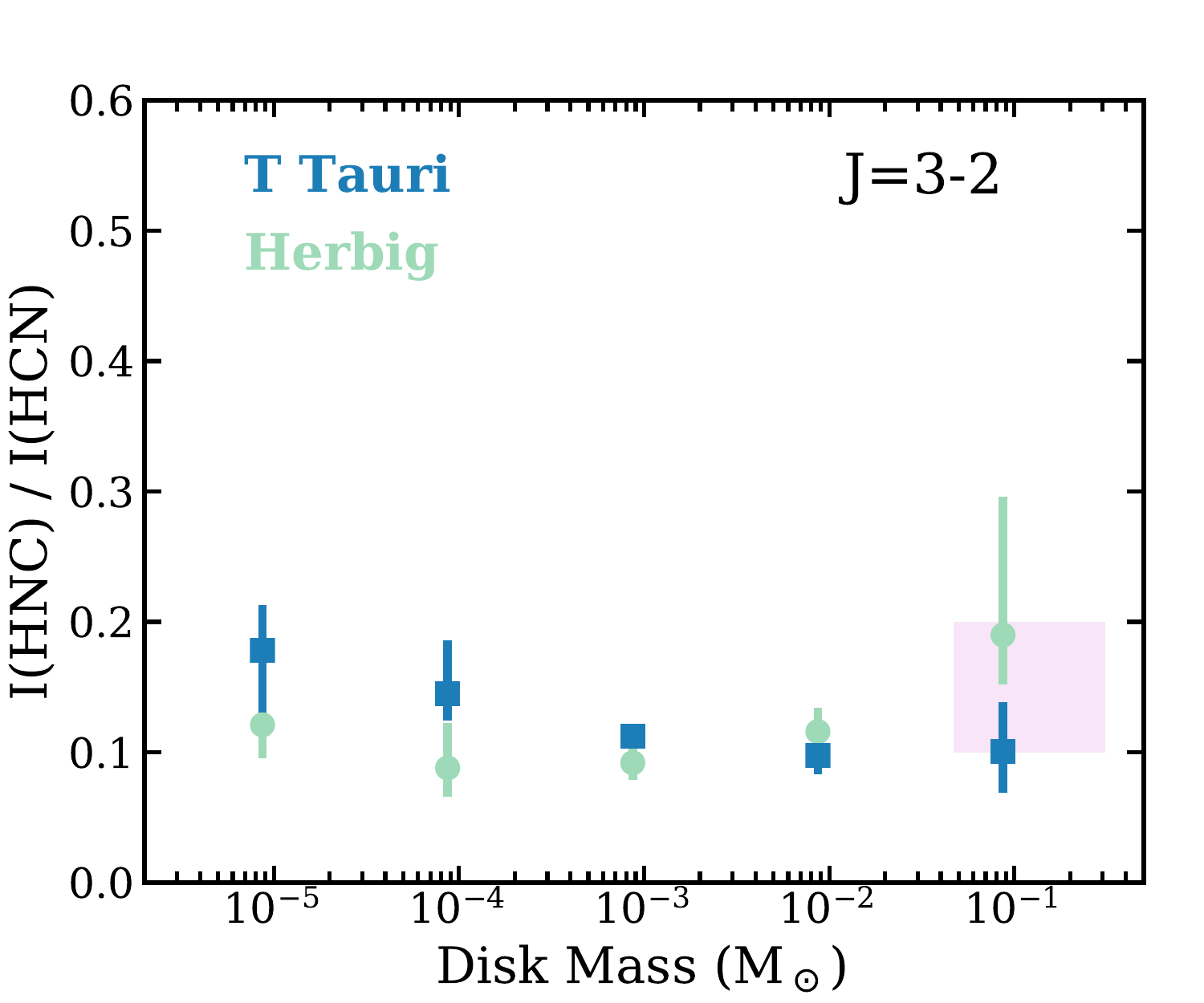} 
\caption{Disk-averaged HNC and HCN line ratio for the T Tauri and Herbig star models and for both transitions. The uncertainties are given by the changes in the vertical disk structure. Observations of TW Hya and HD 163296 are marked for comparison  \citep{Graninger2015}. \label{fig:lineratio_diskaverage}}
\end{figure*}

Disk-integrated HNC-to-HCN line ratios are rather constant in our models with varying stellar and disk conditions, with 0.2$\sim$0.3 for the $1-0$ transition and  0.1$\sim$0.2 for the $3-2$ transition (Figure~\ref{fig:lineratio_diskaverage}). The differences reflect their emitting regions. Our model predictions are consistent with current available observations of TW Hya and HD 163296.

\section{HNC-to-HCN abundance ratio with C and O depletion}
The volatile carbon and oxygen abundances affect the HNC-to-HCN ratio as a disk thermometer because the reactions of HNC with C and O are important pathways for HNC destruction. With enhanced C and O depletion, the HNC-to-HCN abundance ratio increases in the cold midplane regions.

\begin{figure*}[!th]
\centering
    \includegraphics[width=0.3\linewidth]{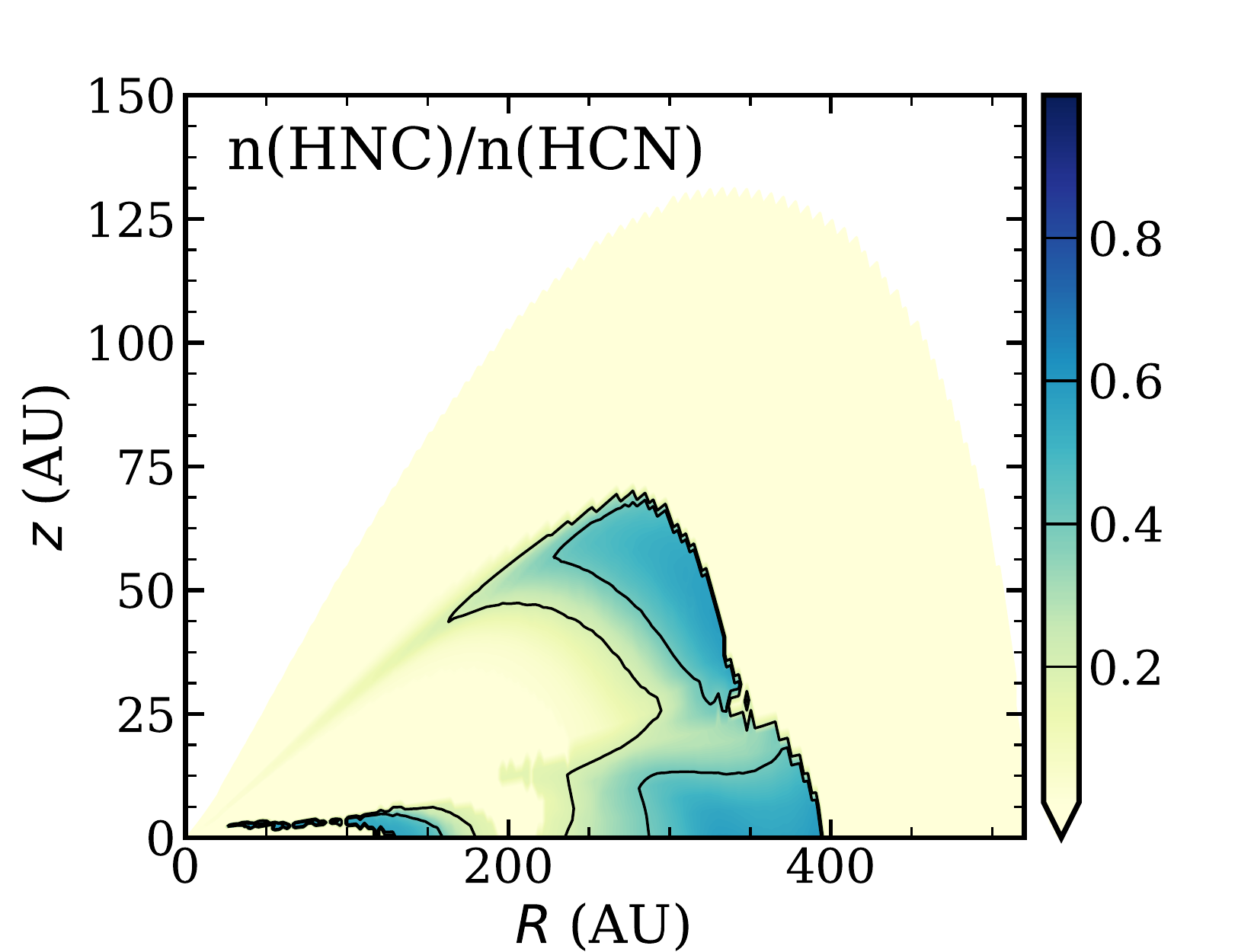}
    \includegraphics[width=0.3\linewidth]{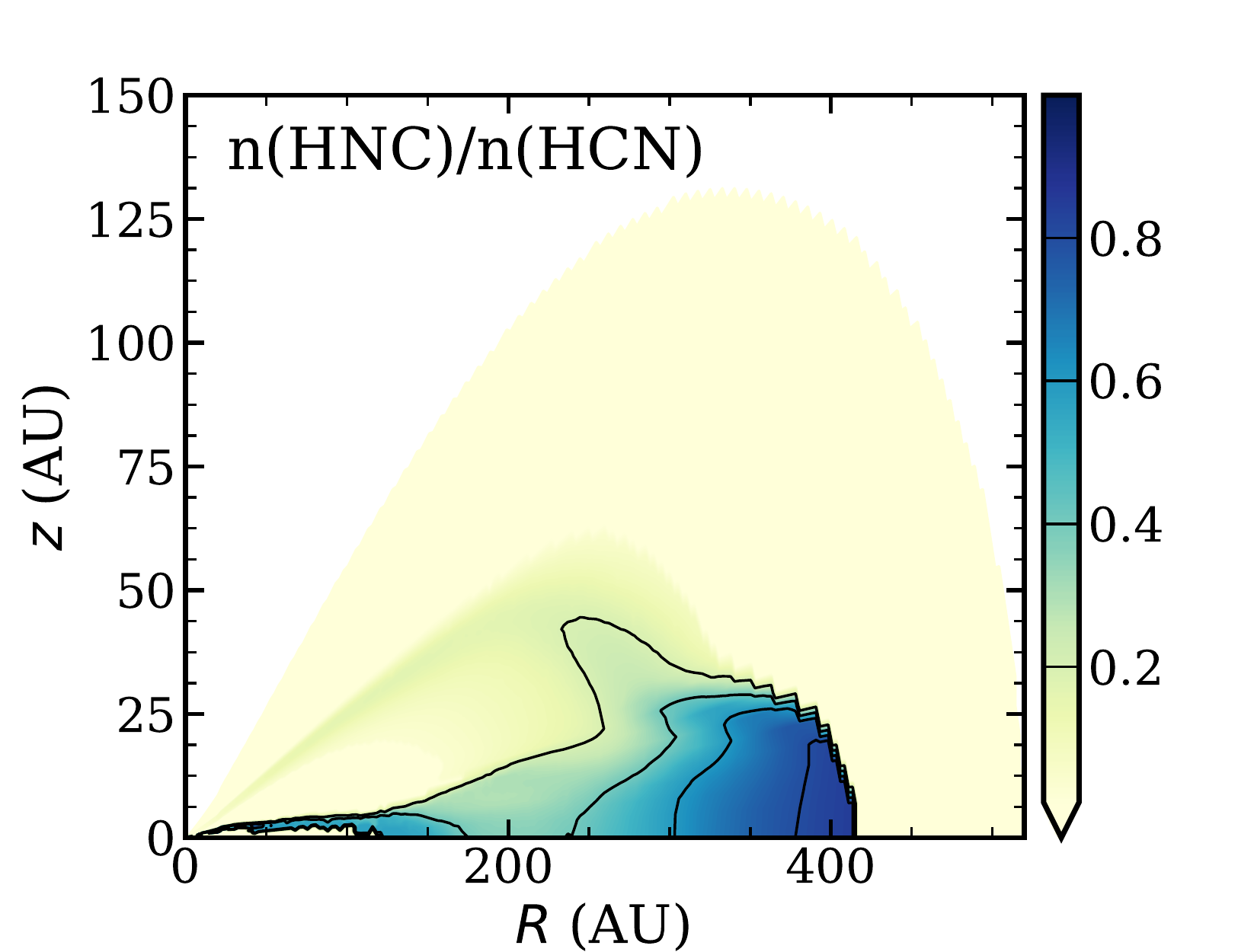} 
    \includegraphics[width=0.3\linewidth]{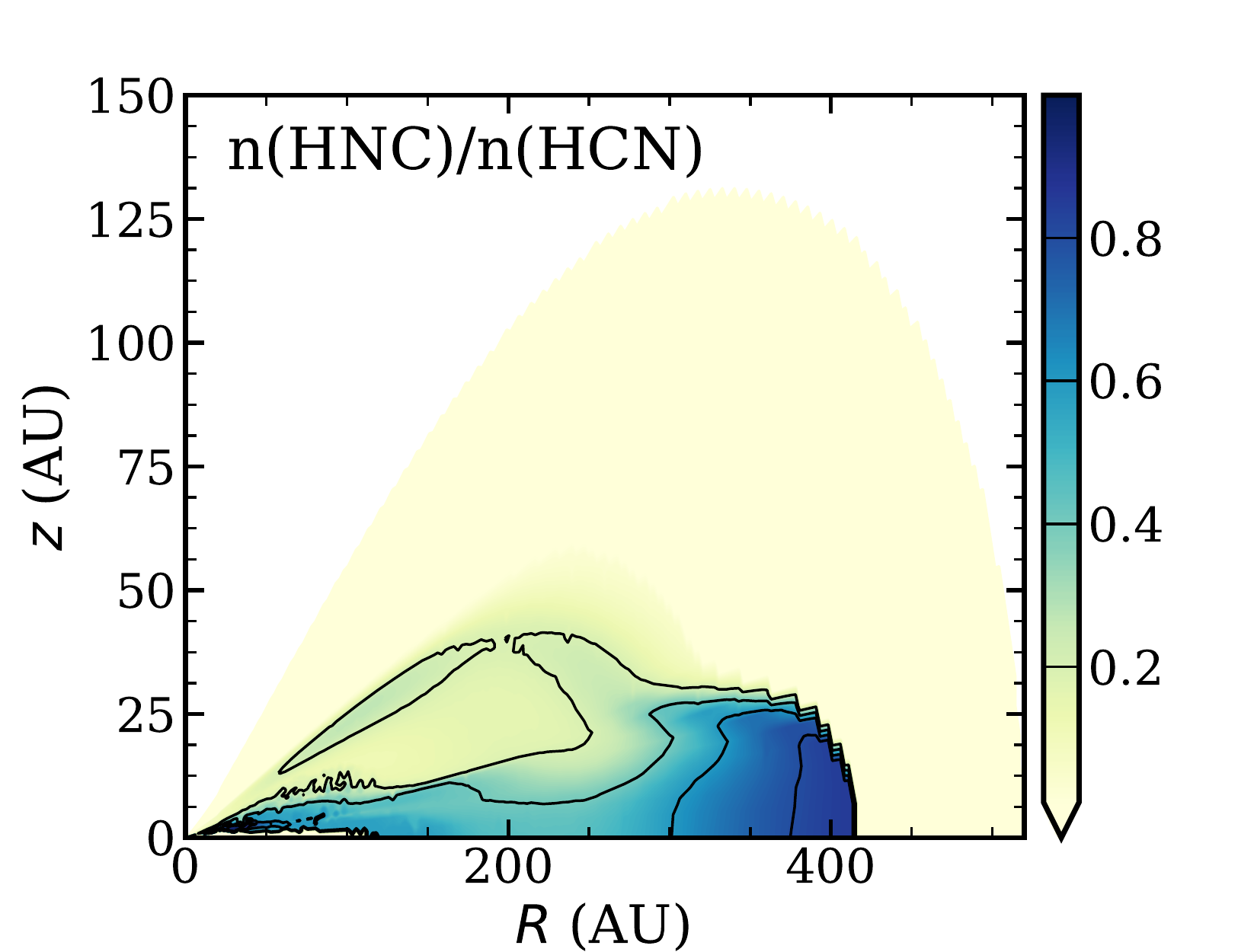} 
\caption{Similar plot as in Figure~\ref{fig:2Dratio}, but for models including initial elemental depletion. From left to right: (1) C and O depleted by a factor of 10, (2) C and O depleted by a factor of 100, and (3) C depleted by a factor of 100, with O further depleted to reach C/O = 1.5.  \label{fig:2Dratio_depletion}}
\end{figure*}

\section{HNC line fluxes of the $2-1$ and $4-3$ transitions}
\begin{figure*}[!t]
\centering
    \includegraphics[width=0.9\linewidth, trim=0 20 0 0]{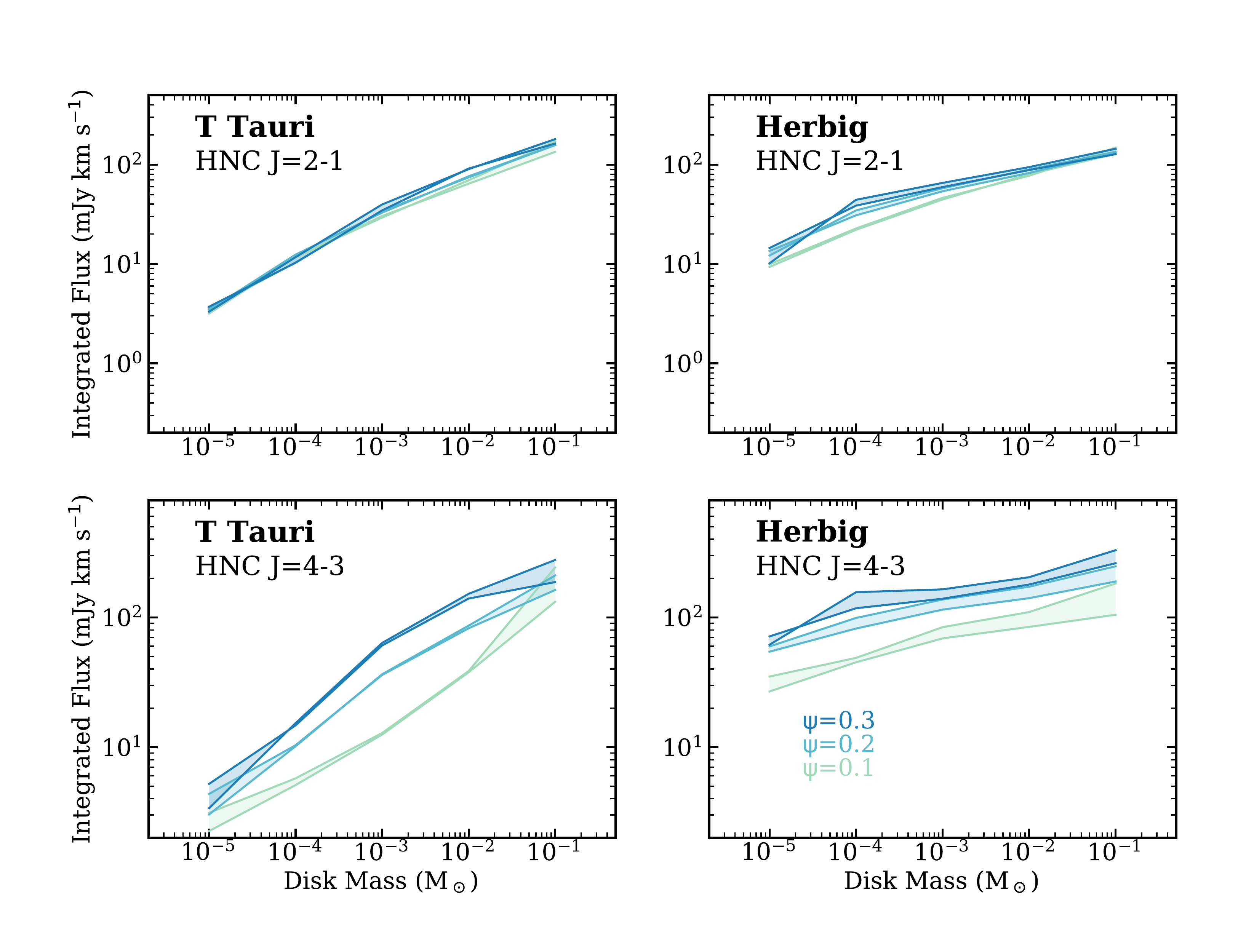}
\caption{Disk-integrated line fluxes (calculated at a distance of 150 pc) of HNC $J=2-1$ and $J=4-3$ as a function of disk mass (left for disks around T Tauri stars, and right for Herbig stars). Colors represent different levels of disk flaring. The upper and lower panels show different transitions.  \label{fig:flux-hnc2143}}
\end{figure*}

Figure~\ref{fig:flux-hnc2143} summarizes the HNC $J=2-1$ and $J=4-3$ line fluxes in our model grids, with similar trends as discussed for the $J=1-0$ and $J=3-2$ lines. Higher-$J$ transitions are more sensitive to UV radiation.

\end{document}